\documentclass[a4paper,aps,pre,reprint,twocolumn,raggedfooter,showpacs]{revtex4-1}
\usepackage{graphicx}
\usepackage[caption=false]{subfig}
\usepackage{latexsym}
\usepackage{amsmath}
\usepackage[makeroom]{cancel}
\usepackage{relsize}
\usepackage{letltxmacro}
\usepackage{amssymb}

\usepackage[usenames]{color}

\newcommand{\sn}[1]{{\color{black}{#1}}}

\LetLtxMacro{\oldsqrt}{\sqrt} 
\renewcommand{\sqrt}[1][\ ]{%
  \def\DHLindex{#1}\mathpalette\DHLhksqrt}
\def\DHLhksqrt#1#2{%
  \setbox0=\hbox{$#1\oldsqrt[\DHLindex]{#2\,}$}\dimen0=\ht0
  \advance\dimen0-0.2\ht0
  \setbox2=\hbox{\vrule height\ht0 depth -\dimen0}%
  {\box0\lower0.71pt\box2}}

\begin{document}

\title{Elastic regimes of sub-isostatic athermal fiber networks}
\author{A. J. Licup}\affiliation{Department of Physics and Astronomy, VU University Amsterdam, The Netherlands}
\author{A. Sharma}\affiliation{Department of Physics and Astronomy, VU University Amsterdam, The Netherlands}
\author{F. C. MacKintosh}\affiliation{Department of Physics and Astronomy, VU University Amsterdam, The Netherlands}
\date{\today}

\begin{abstract}
Athermal models of disordered fibrous networks are highly useful for studying the mechanics of elastic networks composed of stiff biopolymers. The underlying network architecture is a key aspect that can affect the elastic properties of these systems, which include rich linear and nonlinear elasticity. Existing computational approaches have focused on both lattice-based and off-lattice networks obtained from the random placement of rods. It is not obvious, a priori, whether the two architectures have fundamentally similar or different mechanics. If they are different, it is not clear which of these represents a better model for biological networks. Here, we show that both approaches are essentially equivalent for the same network connectivity, provided the networks are sub-isostatic with respect to central force interactions. Moreover, for a given sub-isostatic connectivity, we even find that lattice-based networks in both 2D and 3D exhibit nearly identical nonlinear elastic response. We provide a description of the linear mechanics for both architectures in terms of a scaling function. We also show that the nonlinear regime is dominated by fiber bending and that stiffening originates from the stabilization of sub-isostatic networks by stress. We propose a generalized relation for this regime in terms of the self-generated normal stresses that develop under deformation. Different network architectures have different susceptibilities to the normal stress, but essentially exhibit the same nonlinear mechanics. Such stiffening mechanism has been shown to successfully capture the nonlinear mechanics of collagen networks.
\end{abstract}

\pacs{}
\keywords{}
\maketitle

\section{Introduction}
\label{sec:intro}
The elastic stress response of living cells and tissues is governed by the viscoelasticity of complex networks of filamentous proteins such as the cytoskeleton and the extracellular matrix~\cite{art:JanmeyCellBio,art:wachsstock1994cross,art:Kasza,art:Bausch,art:Fletcher,art:chaudhuri2007reversible, art:tharmann2007viscoelasticity,art:picu2011mechanics,art:BroederszRMP}. This property of such biological gels not only makes living cells and tissues stiff enough to maintain shape and transmit forces under mechanical stress, but also provides them the compliance to alter their morphology needed for cell motion and internal reorganization. Unlike ordinary polymer gels and other materials with rubber-like elastic properties however, biological gels behave nonlinearly in response to deformation. One classic feature is strain stiffening, where a moderately increasing deformation leads to a rapid increase in stress within the material. Such is observed in gels of cytoskeletal and extracellular fibers \cite{art:Gardel,art:janmey1983rheology,art:Xu,art:chaudhuri2007reversible,art:tharmann2007viscoelasticity,art:kabla2007nonlinear, art:Storm,art:didonna2006filamin,art:GardelPNAS06,art:WagnerPNAS06,art:KaszaPRE09, art:stein2011micromechanics,art:picu2011mechanics} and in soft human tissues \cite{art:Saraf}. Another interesting aspect of elastic nonlinearity is the so-called negative normal stress. Most solid materials exhibit what is known as the Poynting effect \cite{art:Poynting} where the response is to expand in a direction normal to an externally applied shear stress. This effect explains why metal wires increase in length under torsional strain. By contrast, cross-linked biopolymer gels exhibit the opposite response to shear deformation, which can be understood either in terms of the inherent asymmetry in the extension-compression response of thermal semiflexible polymers or non-affine deformations in athermal fiber networks \cite{art:Janmey2,art:HeussingerPRE07,art:Conti}.%

Research on the elastic properties of fiber networks often aimed to elucidate the microscopic origins of viscoelasticity has generated significant progress, making way for models that highlight the importance and interplay of semiflexible filaments, cross-link connectivity, network geometry, and disorder. An important consideration when modeling the elastic response of biological gels with fiber networks is the inherent instability of the underlying geometry with respect to stretching. Whether intracellular or extracellular biopolymer networks are studied, the constituent fibers usually form either cross-linked or branched architectures~\cite{art:Lindstrom,art:LicupPNAS,art:Sharma}, corresponding to an average connectivity below the Maxwell isostatic criterion for marginal stability of spring networks with only stretching response. Such systems, however, can be stabilized by a variety of additional interactions, such as fiber bending rigidity \cite{art:BroederszRMP,art:Wyart,art:BroederszNatPhys}, thermal fluctuations \cite{art:DennisonPRL}, internal stresses generated by molecular motors \cite{art:SheinmanPRL,art:shokef2012scaling}, boundary stresses \cite{art:LicupPNAS}, or even strain \cite{art:Sharma,art:Fengarxiv2015}. These stabilizing fields give rise to interesting linear and nonlinear elastic behavior.

Detailed analytical and computational work on the linear elastic response of networked systems reveal two distinct regimes: an affine regime dominated by extension/compression of the fibers and a cross-over to a non-affine one dominated by fiber bending \cite{art:Wilhelm,art:HeadPRL,art:HeadPRE,art:DasPRL2007}. In addition to fiber elasticity, these linear regimes are also found to be dictated by network structure and disorder and can exhibit rich zero-temperature critical behavior, including a cross-over to a mixed regime \cite{art:BroederszNatPhys}. Such linear regimes in turn have important consequences to the nonlinear response where large deformations are involved. In particular, large stresses applied to a network initially dominated by filament bending would lead to a strong strain-induced stiffening response \cite{art:Sharma}, which coincides with the onset of negative normal stress \cite{art:Janmey2,art:Conti}.%

In general, the variety of computational models to understand certain specific aspects of linear or nonlinear network elasticity can either be based on off-lattice \cite{art:Wilhelm,art:HeadPRL,art:HeadPRE,art:shahsavari2012model,art:Conti,art:Onck,art:Huisman} or lattice structures \cite{art:BroederszNatPhys,art:BroederszSM2011,art:BroederszPRL2012,art:Heussinger,art:MaoPRE042602_2013}, which can also be combined with a mean-field approach \cite{art:BroederszNatPhys,art:DasPRL2007,art:SheinmanPRE,art:MaoPRE042601_2013}. Indeed, much has been done with lattices to understand linear elasticity, in contrast to nonlinear elasticity often studied on random networks. The advantage of lattice models is the computational efficiency as well as the relative ease with which one can generate increasingly larger network sizes. We intend to study nonlinear elasticity using a lattice based model and compare with results on a random network. We begin with a detailed description of the disordered phantom network used to study the elastic stress response of passive networks with permanent cross-links \cite{art:huisman2007three,art:BroederszSM2011,art:BroederszPRL2012}. This model allows independent control of filament rigidity, network geometry and cross-link connectivity. We present our results in the nonlinear elastic regime, focusing on shear stiffening and negative normal stress. Finally we conclude with implications when using lattice-based models to understand nonlinear elasticity of stiff fiber networks.%

\section{Modeling sub-isostatic athermal networks}
\label{sec:model}
Biopolymers can form either cross-linked or branched network structures that have average connectivity somewhere between three-fold ($z=3$) at branch points and four-fold ($z=4$) at cross-links~\cite{art:Lindstrom,art:LicupPNAS,art:Sharma}. If these nodes interact only via central forces such as tension or compression of springs, the network rigidity vanishes and and the resulting networks are inherently unstable~\cite{art:Maxwell}. However, it is known that these sub-isostatic systems can be stabilized by other effects such as the bending of rigid fibers \cite{art:HeadPRL,art:Alexander,art:Wyart,art:BroederszNatPhys}. In this section, we describe a minimal model of a sub-isostatic network in which the the constituent fibers are modeled as an elastic beam whose rigidity is governed by pure enthalpic contributions.

\subsection{Network generation}
\label{subsec:networkmodel}
We generate a disordered phantom network~\cite{art:BroederszSM2011,art:BroederszPRL2012} by arranging fibers into a $d$-dimensional space-filling regular lattice of size $W^d$ \sn{(no. of nodes)}. \sn{We use triangular and FCC lattices for $d=2$ and $d=3$, respectively.  The network occupies a total volume (or area for 2D lattices) $V=v_0 W^d$, where $v_0$ is the volume (or area) of a unit cell.} Periodic boundaries are imposed to reduce edge effects. Freely-hinged cross-links bind the intersections of fiber segments permanently at the vertices, which are separated by a uniform spacing $\ell_0$. Since a full lattice has a fixed connectivity of either $z_\mathrm{max}=6$ (2D) or $z_\mathrm{max}=12$ (3D), we randomly detach binary cross-links (i.e., $z=4$) at each vertex. \sn{Starting from a 2D triangular network, this results in an average distance $l_c$ between cross-links, where $l_c=3\ell_0/2$, while $l_c=\ell_0$ for the 3D FCC lattice. In either case, this procedure creates a network with connectivity $z=4$ composed of \emph{phantom} segments that can move freely and do not interact with other segments, except at cross-links.} Thus far, all fibers span the system size which leads to unphysical stretching contributions to the macroscopic elasticity. We therefore cut at least one bond on each spanning fiber. Finally, to reduce the average connectivity \sn{to physical values of $z<4$,} we \emph{dilute} the lattice by cutting random bonds with probability \sn{$q=1-p$, where $p$ is the probability of an existing bond.} Any remaining dangling ends are further removed. The lattice-based network thus generated is sub-isostatic with average connectivity $2<z<4$, average fiber length $L=\ell_0/q$ and average distance between cross-links \sn{$l_c=\ell_0$ for an initial undeformed FCC lattice and $l_c\simeq1.4\ell_0$ for an intial triangular lattice with $z\simeq3.2$.}

\sn{Mikado networks are generated by random deposition of monodisperse fibers of unit length onto a 2D box with an area $W\times W$. A freely-hinged cross-link is inserted at every point of intersection resulting in a local connectivity of $4$. However, some of the local bonds are dangling ends and are removed from the network thus bringing the average connectivity below $4$. The deposition continues until the desired average connectivity is obtained.}

\sn{For the rest of this work, we use $l_c$ to denote the average distance between crosslinks for both lattice-based and Mikado networks. For simplicity and unless otherwise stated, we use $l_c=\ell_0$ for both 2D and 3D lattice-based networks.}

\subsection{Fiber elasticity}
\label{subsec:fiber_elast}
\sn{In modeling fiber networks, each fiber can be considered as an Euler-Bernoulli or Timoshenko beam~\cite{art:huisman2007three,art:shahsavari2012model}. From a biological perspective, it is important to consider the semiflexible nature of the fibers to account for the finite resistance to both tension and bending.}
When the network is deformed, any point on every fiber undergoes a displacement which induces a local \sn{fractional} change in length $\frac{dl}{ds}$ and a local curvature $\big|\frac{d\hat{t}}{ds}\big|$. The elastic energy thus stored in the fiber is given by~\cite{art:HeadPRE}
\begin{equation}
\label{eqn:eWLC}
\mathcal{H}_f=\frac{\mu}{2}\int_f\left(\frac{dl}{ds}\right)^2ds+\frac{\kappa}{2}\int_f\left|\frac{d\hat{t}}{ds}\right|^2ds,
\end{equation}
\sn{where the parameters $\mu$ and $\kappa$ describe the 1D Young's (stretch) modulus and bending modulus, respectively. The integration} is evaluated along the \sn{undistorted} fiber contour. \sn{The total energy $\mathcal{H}=\sum_f\mathcal{H}_f$ is the sum of Eq.~(\ref{eqn:eWLC}) over all fibers.}

Treating the fiber as a homogeneous cylindrical elastic rod of radius $a$ and Young's modulus $E$, we have from classical beam theory~\cite{book:LandauLifshitz} $\mu=\pi a^2 E$ and $\kappa=\tfrac{1}{4}\pi a^4 E$. These parameters can be absorbed into a \emph{bending length scale} $l_b=\sqrt{\kappa/\mu}=a/2$. One can normalize $l_b$ by the geometric length $l_c$ to obtain a dimensionless fiber rigidity $\tilde\kappa=(l_b/l_c)^2$, or
\begin{equation}
\label{eqn:dimlesskappa}
\tilde\kappa=\frac{\kappa}{\mu l_c^2}.
\end{equation}
\sn{As noted in Sec.~\ref{subsec:networkmodel}, for simplicity we take $l_c$ to be the lattice spacing $\ell_0$ of the 2D and 3D lattice-based networks. For Mikado networks, $l_c$ is the average spacing between crosslinks.}

In our network of straight fibers with discrete segments, a midpoint node is introduced on every segment to capture at least the first bending mode over the smallest length scale $l_c$. The set of spatial coordinates $\{r_j\}$ of all nodes (i.e., cross-links, phantom nodes and midpoints) thus constitutes the internal degrees of freedom of the network. Under any macroscopic deformation, e.g. \emph{simple shear} strain $\gamma$, the nodes undergo a displacement $\{r_j\}\rightarrow\{r'_j\}$ which induces the dimensionless local deformations $\lambda_j=\delta\ell_j/\ell_j$ and $\theta_j=|\hat{t}_{j,j+1}-\hat{t}_{j-1,j}|$. Here, $\delta\ell_j=\ell'_j-\ell_j$ is the length change of a fiber segment with rest length $\ell_{j}=|r_{j+1}-r_j|$ and $\hat{t}_{i,j}$ is a unit vector tangent to segment $\langle ij\rangle$. The fiber then stores an elastic energy expressed as a discretized form of Eq.~\eqref{eqn:eWLC}:
\[
\mathcal{H}_f=\frac{1}{2}\sum_{j\in f}\left(\mu\ell_{j}\lambda_{j}^2+\frac{\kappa}{l_j}\theta_j^2\right),
\]
where $l_j=\frac{1}{2}\left(\ell_{j-1}+\ell_{j}\right)$. By taking $l_j\simeq\ell_j\simeq l_c$, we can rewrite this equation with an explicit dependence on deformation and fiber rigidity as:
\begin{equation}
\label{eqn:Efiber}
\mathcal{H}_f\left(\gamma,\tilde{\kappa}\right)=\mu l_c\sum_{j\in f}\hat{\mathcal{H}}_j\left(\gamma,\tilde{\kappa}\right),
\end{equation}
where $\hat{\mathcal{H}}_j=\frac{1}{2}(\lambda_{j}^2+\tilde{\kappa}\theta_j^2)$ is a dimensionless elastic energy of a fiber segment. Note that the dependence on $\{\lambda_j,\theta_j\}$ is accounted for by the macroscopic \sn{strain} $\gamma$.

\subsection{Network Elasticity}
\label{subsec:net_elast}
The network elasticity is determined not only by the rigidity of the constituent fibers \sn{but also by the network connectivity, which we characterize equivalently by $z$ or the average \emph{cross-linking density} $L/l_c$, that is also the number of cross-links per fiber. This ratio has been shown to govern the network's affine/non-affine response to the applied deformation \cite{art:HeadPRL,art:BroederszPRL2012}.  A higher density of cross-links leads to a more affine (i.e., uniform) deformation field. By contrast, fewer cross-links per fiber allows the possibility of exploring non-uniform displacements resulting in a non-affine response \cite{art:Wilhelm,art:Heussinger}.} Effectively, the network elasticity can be characterized by $\tilde\kappa$ and $L/l_c$. 

\sn{The stress and moduli depend on the energy density $\mathcal{U}$, i.e., energy per unit volume. Since the expression for the total energy involves an integral along the contour length of all fibers, $\mathcal{U}$ is naturally proportional to the total length of fiber per unit volume. Thus, $\rho$, together with the energy per length, $\mu$, set the natural scale for energy density, stress and modulus. Thus, we write
\begin{equation}
\label{eqn:energydensity}
\mathcal{U}=\mu\rho\langle\hat{\mathcal{H}}_j\left(\gamma,\tilde{\kappa}\right)\rangle_\mathrm{s}=\frac{\mu}{l_c^{d-1}}\tilde{\mathcal{U}}(\gamma,\tilde{\kappa}),
\end{equation}
where $\langle\cdot\rangle_\mathrm{s}$ is an average over all fiber segments.} Expressing $\rho$ as $\tilde{\rho}l_c^{1-d}$ where $\tilde{\rho}$ is a dimensionless number of fiber segments in a unit volume, we have $\tilde{\mathcal{U}}=\tilde{\rho}\langle\tilde{\mathcal{H}}_j\left(\gamma,\tilde{\kappa}\right)\rangle_\mathrm{s}$. Successively differentiating  Eq.~\eqref{eqn:energydensity} with respect to $\gamma$, one obtains $\sigma=\tfrac{\partial\mathcal{U}}{\partial\gamma}=\mu\rho\tilde{\sigma}(\gamma,\tilde{\kappa})$ and $K=\tfrac{\partial\sigma}{\partial\gamma}=\mu\rho\tilde{K}(\gamma,\tilde{\kappa})$.

In our simulations, the line density $\rho$ is specific to the chosen network architecture. In the lattice-based networks, we have $\tilde{\rho}_\mathrm{2D}=\tfrac{6p}{\oldsqrt{3}}$ and $\tilde{\rho}_\mathrm{3D}=\tfrac{12p}{\oldsqrt{2}}$ (see Appendix). \sn{With $l_c = l_0$ in lattice-based networks, the line density can be easily calculated for any given bond dilution probability $q$ (See Appendix). For the off-lattice Mikado network, one can also define an average distance $l_c$ between crosslinks. However, one does not need to know $l_c$ explicitly to calculate the line density $\rho_\mathrm{M}$ of a Mikado network: $\rho_\mathrm{M}=\tilde{\rho}_\mathrm{M}/L$, where $\tilde{\rho}_\mathrm{M}=n L^2$ and $n$ is the number of fibers per unit area~\cite{art:HeadPRE_R}. The line density $\rho$ is thus explicitly known for lattice and off-lattice models and as we show below, can be used to draw a quantitative comparison between the two computational approaches.} \sn{It also follows that} comparison between simulation results and experiments is possible by accounting for the line density $\tilde{\rho}$ of the specific network architecture. In particular, any measured quantity $X$ (e.g. stress or modulus) must be compared as $\tfrac{X}{\mu\l_c^{1-d}}=\tilde{\rho}\tilde{X}(\gamma,\tilde{\kappa})$, or as \sn{$\tfrac{X}{\mu L^{1-d}}=\tilde{\rho}_\mathrm{M}\tilde{X}(\gamma,\tilde{\kappa})$} in the case of Mikado networks. \sn{Since $\tilde{\kappa}$ is dimensionless, different network architectures for a fixed connectivity $z$ can be characterized by their respective $\tilde{\rho}$.}

\sn{For 3D networks, the dimensionless fiber rigidity $\tilde{\kappa}$ is also related to the material concentration in a biopolymer network through the volume fraction of rods. For any given network structure of stiff rods, a \sn{cylindrical} segment of length $l_c$ and cross-section $\pi a^2$ occupies a volume fraction $\phi=\pi a^2\rho\propto a^2/ l_c^2$. Since the fiber rigidity $\tilde\kappa=\kappa/\mu l_c^2\sim a^2/l_c^2$, we obtain $\tilde\kappa\propto\phi$.} Indeed, it has been shown that reconstituted collagen network mechanics \sn{is} consistent with a reduced fiber rigidity $\tilde{\kappa}$ that is proportional to the protein concentration~\cite{art:LicupPNAS,art:Sharma}.
\begin{figure}[t]
\centering
\includegraphics[width=0.48\textwidth]{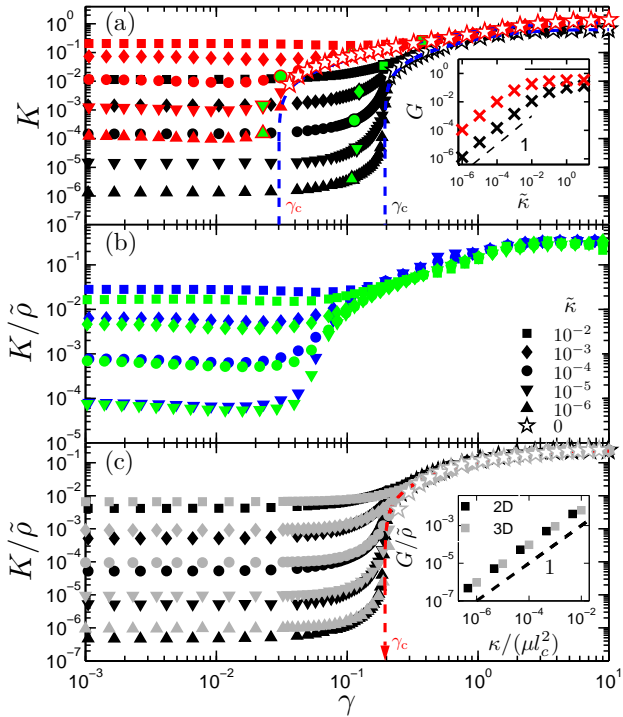}
\caption{(Color online) (a) Stiffness $K$ of a 2D lattice-based network as a function of the macroscopic strain $\gamma$. The black data corresponds to $L/l_c=3$ ($z=3.2$) while the red data is for $L/l_c=9$ ($z=3.8$). In both cases, the stiffness is constant for low $\gamma$. At the onset of nonlinear stiffening marked by green symbols, $K$ increases rapidly until $\gamma=\gamma_\mathrm{c}$, defined in the limit of $\tilde{\kappa}=0$ (blue dashed curves). For $\gamma\gg\gamma_\mathrm{c}$, all curves collapse as stiffening is independent of $\tilde{\kappa}$ and dominated by fiber stretching. The \sn{strain $\gamma_0$ at the} onset of nonlinearity shifts to lower strains with increasing $L/l_c$. Inset: The linear modulus $G$ plotted as a function of fiber rigidity $\tilde{\kappa}$ also shows two elastic regimes: $G\sim\tilde{\kappa}$ (dashed line of unit slope) and $G\sim\tilde{\kappa}^0$ (solid horizontal line). Symbol colors represent the same $L/l_c$ values in the main plot. (b) 2D Mikado (green data, $L/l_c=11$, $z=3.6$) and 2D \sn{lattice-based} (blue data, $L/l_c=6$, $z=3.6$) \sn{network} simulations normalized by their respective $\tilde{\rho}$, show the same qualitative behavior. (c) Stiffening curves from 3D (gray data) and 2D (black data) \sn{lattice-based networks}, both with $z=3.2$ show the same qualitative behavior as well as the same $\gamma_\mathrm{c}$. The 3D and 2D data are each normalized by $\tilde{\rho}_\mathrm{3D}$ and $\tilde{\rho}_\mathrm{2D}$. \sn{Inset: For the same $z=3.2$, the normalized linear modulus $G/\tilde{\rho}$ in 2D networks become virtually indistinguishable from 3D when plotted versus $\kappa/\mu l_c^2$, using the average distance $l_c$ between crosslinks, i.e., $l_c\simeq 1.4\ell_0$ in 2D and $l_c=\ell_0$ in 3D.}}
\label{fig:Kvgamma}
\end{figure}

To explore the elastic response of the network, the volume-preserving \sn{simple shear} strain $\gamma$ is increased in steps over a range that covers all elastic regimes, typically from $0.1\%$ to $1000\%$. At each $\delta\gamma$ strain step, the total elastic energy density is minimized by relaxing the internal degrees of freedom using a conjugate gradient minimization routine~\cite{book:numrec}. Lees-Edwards boundary conditions~\cite{art:LeesEdwards} ensure that the lengths of segments crossing the system boundaries are calculated correctly. From the minimized total elastic energy density, the shear stress $\sigma$ and differential shear modulus $K$ are evaluated. We also determine the normal stress $\tau=\left.\tfrac{\partial\mathcal{U}}{\partial\varepsilon}\right|_\gamma$ where $\varepsilon$ is a small uniform deformation applied normal to the shear boundaries. Measuring these quantities allows us to characterize the elastic regimes of the network which depends on the rigidity of the constituent fibers, the average density of cross-links, as well as the applied deformation.

One can immediately identify different elastic regimes from the stiffening curves in Fig.~\ref{fig:Kvgamma}a: (i) a linear regime at low strain for which $K=G$ is constant; and (ii) a nonlinear regime showing a rapid increase of $K$ for $\gamma\gtrsim\gamma_0$ where $\gamma_0$ is the strain at the \emph{onset} of nonlinearity. \sn{For networks with longer fibers and higher $L/l_c$, the strain $\gamma_0$ shifts to lower values.} The linear modulus $G$ reveals two distinct regimes as shown in the inset: (1) a bend-dominated regime with $G\sim\tilde{\kappa}$, and (2) a stretch-dominated regime at high $\tilde\kappa$, where bending is suppressed and the response is primarily due to stretching, i.e., $G\sim\mu$. Finally for large strains $\gamma\geq\gamma_\mathrm{c}$, which is the \emph{critical} strain for which a fully floppy $\kappa=0$ network develops rigidity, the stiffness grows independently of $\tilde{\kappa}$ as stretching modes become dominant. Here, the stiffening curves converge to that of the $\kappa=0$ limit. This convergence is indicative of the ultimate dominance of stretching modes over bending for strains above $\gamma_\mathrm{c}$ (see Sec.\ \ref{sec:nonlinregime}).

Interestingly, we find that the characteristic features of stiffening are remarkably insensitive to local geometry (i.e., Mikado vs lattice-based) and even dimensionality, for networks with the same average connectivity $z$. This holds, however, only below the respective isostatic thresholds, which are different in 2D and 3D. Specifically, we show in Fig.~\ref{fig:Kvgamma}b that 2D Mikado and 2D lattice-based networks of the same $z$ show even quantitative agreement, once we account for the difference in fiber density $\mu\tilde{\rho}$. \sn{By simply rescaling the stiffness with $\tilde{\rho}$, it seems that any explicit dependence of stiffness on the local geometry is factored out.} Figure~\ref{fig:Kvgamma}c shows a similar insensitivity to dimensionality, again accounting for network density $\tilde\rho$. \sn{This is even more apparent when plotting the normalized linear modulus $G/\tilde{\rho}$ versus $\kappa/\mu l_c^2$ with the actual $l_c$ for 2D and 3D lattices, as shown in the inset to Fig.~\ref{fig:Kvgamma}c. As noted in Sec.~(\ref{subsec:networkmodel}), we defined the reduced bending rigidity $\tilde\kappa=\kappa/(\mu\ell_0^2)$ for lattice-based networks, although the average distance $l_c$ between crosslinks is somewhat larger than the lattice spacing $\ell_0$ by the construction of our 2D lattice-based networks. Taking the actual values of $l_c$ for 2D ($\simeq1.4\ell_0$) and 3D ($\ell_0$) networks at the same $z=3.2$, one obtains an almost perfect collapse of the data.} Moreover for the same connectivity \sn{($<4$),} even the strain thresholds $\gamma_0$ \sn{and $\gamma_\mathrm{c}$} agree between Mikado and \sn{2D lattice-based networks}, and between 2D and 3D lattice-based networks~\cite{art:LicupPNAS,art:Sharma}.

\section{Linear Regime}
\label{sec:linregime}
\begin{figure*}[t]
\subfloat[]{
  \includegraphics[width=0.47\textwidth]{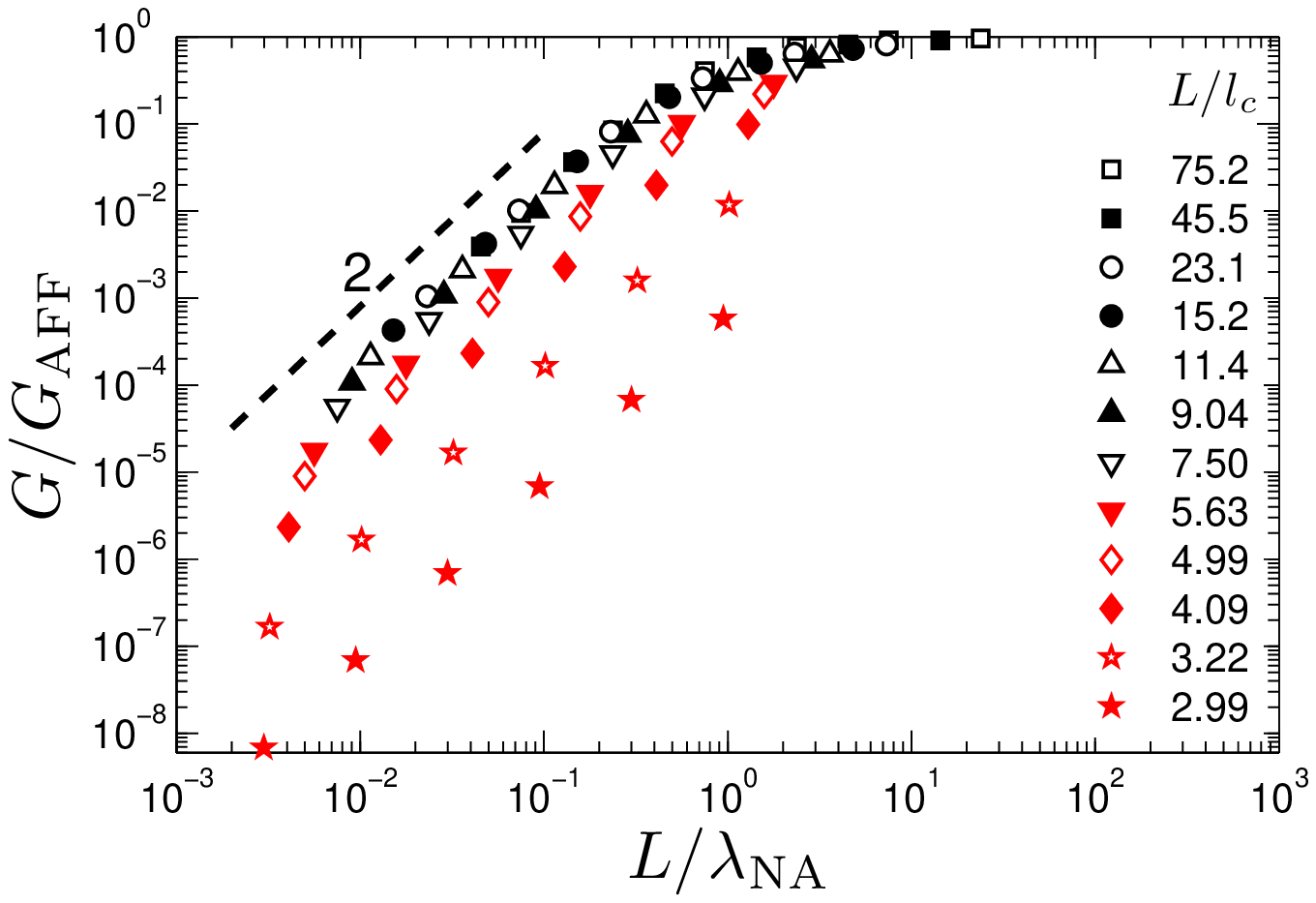}
  \label{fig:nonaffinity}
}
\qquad
\subfloat[]{
  \includegraphics[width=0.47\textwidth]{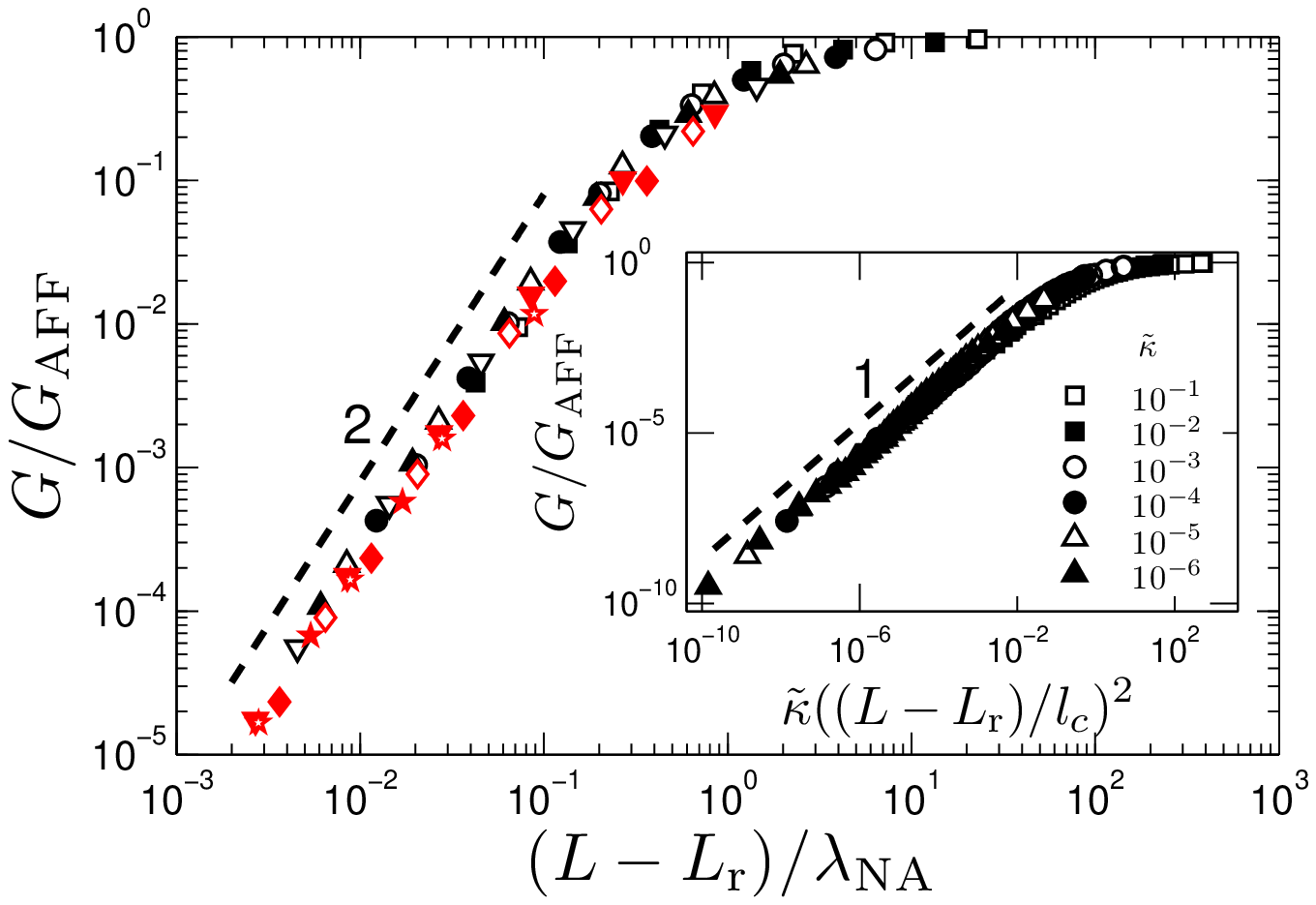}
  \label{fig:nonaffinity_corr}
}
\caption{(Color online) Collapse of linear modulus with non-affinity length scale (a) without and (b) with $L_\mathrm{r}$ correction. Red symbols represent networks in the vicinity of the rigidity percolation regime. The inset of (b) shows the collapse of the linear modulus with $\tilde{\kappa}(L-L_\mathrm{r})$.}
\end{figure*}

The linear regime is characterized by a constant modulus $G$ over the range of $\gamma<\gamma_0$. As mentioned above and shown in the inset of Fig.~\ref{fig:Kvgamma}a, the linear modulus exhibits two distinct regimes: a bend-dominated one in which $G \sim \tilde{\kappa}$ and one in which $G$ is independent of $\tilde{\kappa}$ and is a stretch-dominated regime where $G\sim\mu$. The crossover between the two regimes \sn{has been shown to be governed} by a non-affinity length scale $\lambda_\mathrm{NA}$, which is determined by $l_c$, $l_b$ as follows~\cite{art:HeadPRL,art:HeadPRE,art:BroederszPRL2012}.
\begin{equation}
\lambda_\mathrm{NA} = l_c \left(\frac{l_c}{l_b}\right)^\zeta.\label{eqn:zeta}
\end{equation}
The exponent $\zeta$ depends on the network structure \sn{and the} ratio $L/\lambda_\mathrm{NA}$ determines the crossover between the elastic regimes as
\begin{equation}
\frac{G}{G_\mathrm{AFF}} \sim \left(\frac{L}{\lambda_\mathrm{NA}}\right)^{2/\zeta},
\label{eqn:linmodscaling}
\end{equation}
where $G_\mathrm{AFF}$ is the modulus in the affine limit. In our lattice-based networks, $G_\mathrm{AFF}\sim\mu\ell_0^{1-d}$. For $\lambda_\mathrm{NA} \geq L$, the modulus is governed by bending modes in the network. On the other hand for $\lambda_\mathrm{NA} < L$, the modulus is governed by \sn{stretching modes.}

\sn{Using mean-field arguments, Ref.~\cite{art:HeadPRE} found that $\zeta\simeq2/5$ for off-lattice 2D Mikado networks, while for 3D FCC lattice-based networks, Ref.~\cite{art:BroederszPRL2012} found that $\zeta=1$.
Here, we focus on 2D lattice-based networks and show that $\zeta=1$, as for the 3D FCC-based networks in Ref.~\cite{art:BroederszPRL2012}.} In Fig.~\ref{fig:nonaffinity}, we show $G/G_\mathrm{AFF}$ vs.\ $L/\lambda_\mathrm{NA}$. As can be seen, data obtained for different values of $L/l_c$ collapse on a master curve with slope $2/\zeta=2$. Significant deviation from the master curve is seen for data corresponding to relatively small values of $L/l_c$. This has been observed in a previous study on 3D FCC networks where such is attributed to a different scaling for networks in the vicinity of the rigidity percolation regime~\cite{art:BroederszPRL2012}. However, on replacing $L$ by $(L-L_\mathrm{r})$, where $L_\mathrm{r}\approx 2.94$ is the average fiber length at rigidity percolation, we obtain an excellent collapse for all values of $L/l_c$ with slope $2/\zeta=2$ (Fig.~\ref{fig:nonaffinity_corr}). It follows from \sn{the} above \sn{correction} that in the linear regime $G/G_\mathrm{AFF} \sim \kappa (L-L_\mathrm{r})^2$ as shown in the inset of Fig.~\ref{fig:nonaffinity_corr}. The scaling $G/G_\mathrm{AFF} \sim \kappa L^2$ is known for 3D FCC lattice-based networks for $L \gg L_\mathrm{r}$~\cite{art:BroederszPRL2012}. Interestingly, such scaling behavior has been observed in experiments on hydrogels~\cite{art:Jaspers}. As we show above, the same scaling holds in 2D lattice-based networks.
\begin{figure*}[t]
\subfloat[]{
  \includegraphics[width=0.47\textwidth]{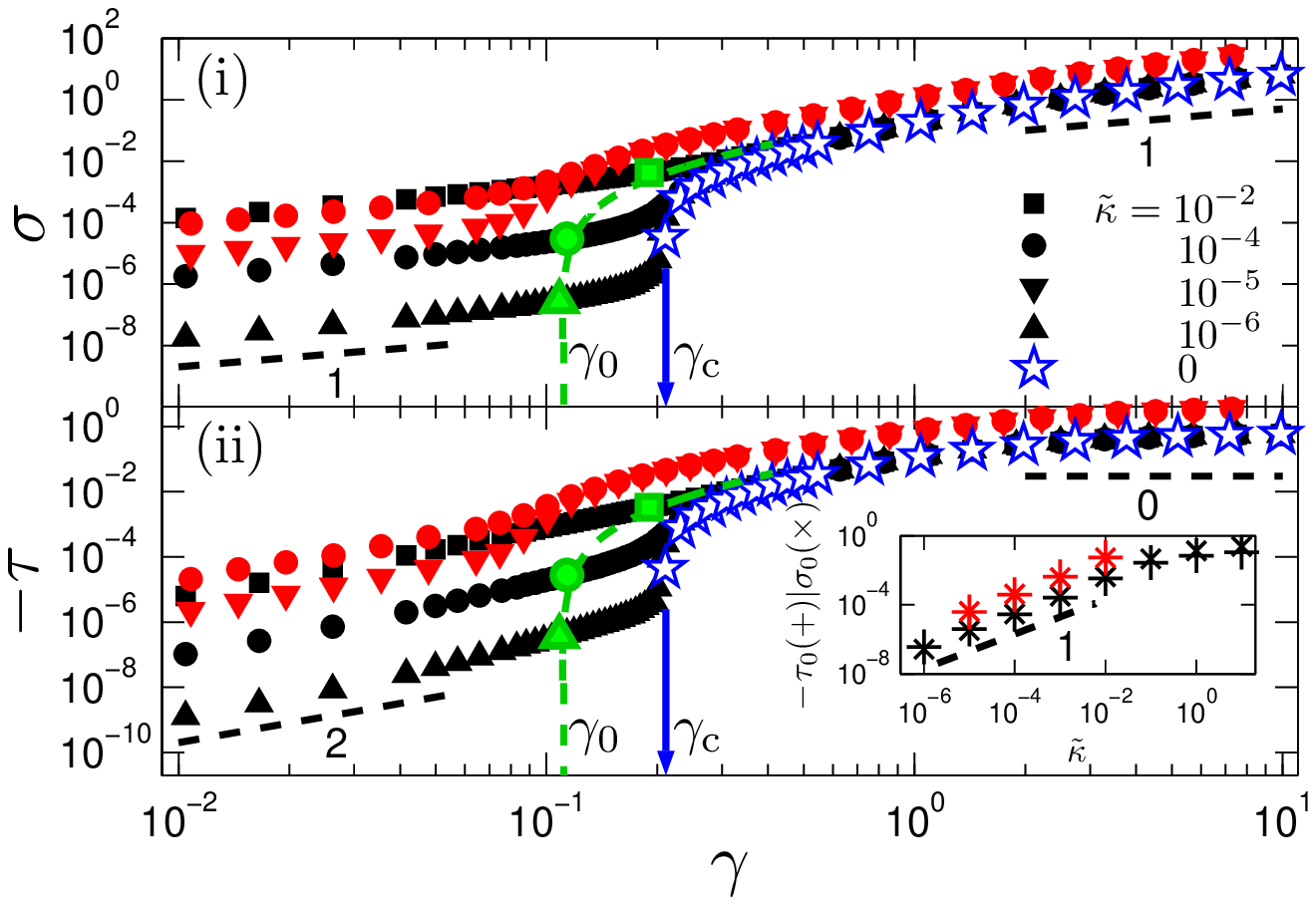}
  \label{fig:Svgamma}
}
\qquad
\subfloat[]{
  \includegraphics[width=0.47\textwidth]{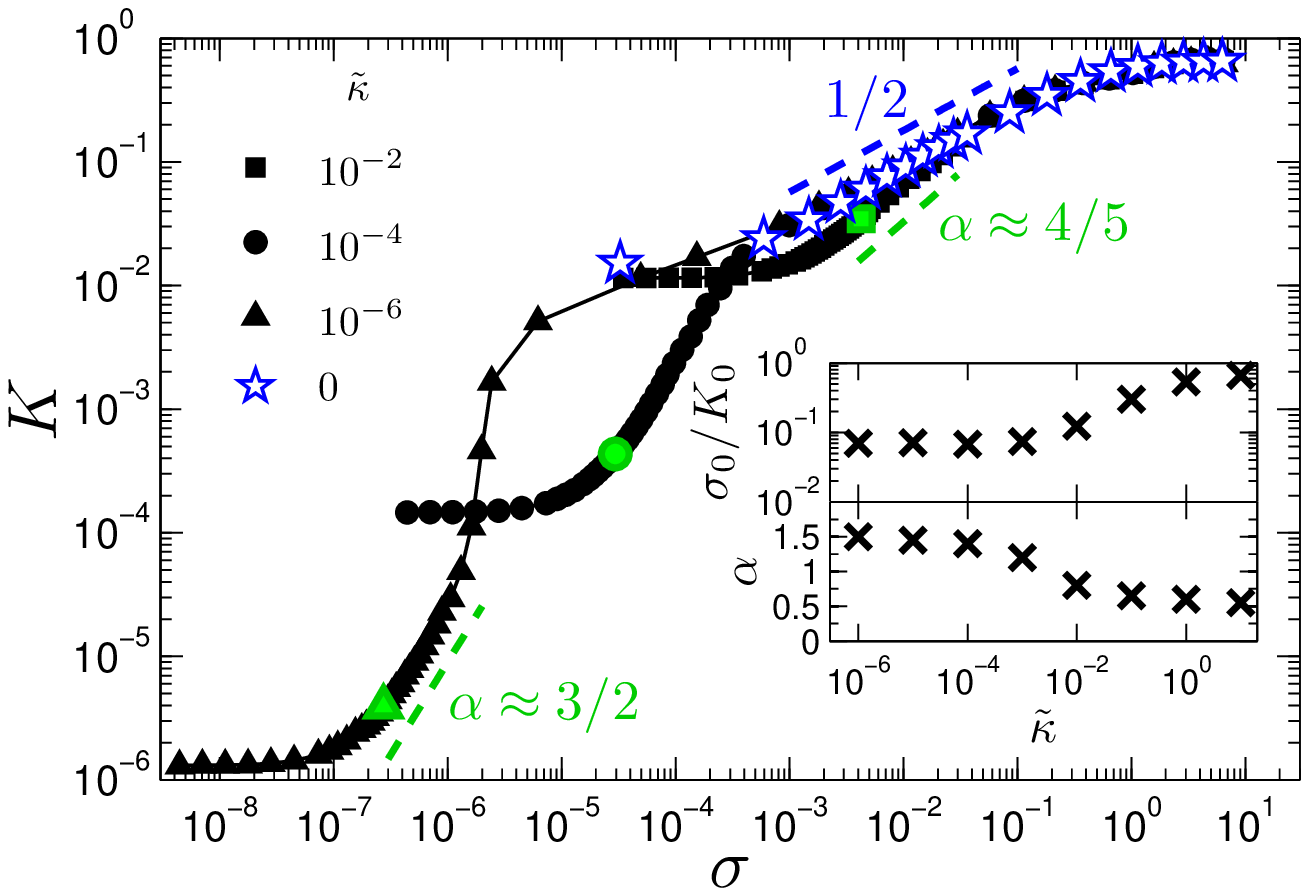}
  \label{fig:KvS}
}
\caption{((Color online) (a) Shear stress $\sigma$ (i) and negative normal stress $-\tau$ (ii) as a function of $\gamma$ and $\tilde{\kappa}$ in a 2D lattice with $L/l_c=3$, $z=3.2$. In the linear regime, $|\sigma|\sim\gamma$ and $|\tau|\sim\gamma^2$. The stresses at the onset strain $\gamma_0$ of stiffening are indicated by green symbols interpolated by the green dashed schematic curve. The blue arrow marks the critical strain $\gamma_\mathrm{c}$. Red data is from the Mikado simulation with $L/l_c=11$, $z=3.6$. Inset: At the onset of stiffening, $-\tau_0\approx\sigma_0$, where both scale linearly with $\tilde{\kappa}$. (b) Stiffness $K$ as a function of shear stress $\sigma$ for different $\tilde{\kappa}$ in the 2D lattice. The lines connecting the data points only serve as visual guides. The green points correspond to $(\sigma_0,K_0)$ at $\gamma_0$ and are replotted in the inset (upper panel) for all $\tilde{\kappa}$. Networks first undergo $K\sim\sigma^\alpha$ stiffening (green dashed lines) followed by $K\sim\sigma^{1/2}$ (blue dashed line). The lower panel of the inset shows the evolution of the stiffening exponent $\alpha$ with fiber rigidity.}
\end{figure*}

With $\zeta=2/5$, the modulus of off-lattice Mikado networks can be quantitatively captured by Eq.~\eqref{eqn:linmodscaling}~\cite{art:HeadPRE,art:HeadPRL}. The mean-field argument implicitly assumes that the non-affinity length scale is larger than the bending correlation length which is given by
\begin{equation}
\lambda_\mathrm{b} = l_c\left(\frac{l_b}{l_c}\right)^\zeta.
\label{eqn:bendingcorrlength}
\end{equation}
Moreover, both $\lambda_\mathrm{NA}$ and $\lambda_\mathrm{b}$ are assumed to be larger than $l_c$. It has been previously pointed out that in the limit of very flexible rods or for low concentrations, Eq.~\eqref{eqn:bendingcorrlength} would predict $\lambda_\mathrm{b} < l_c$, which is an unphysical result~\cite{art:HeadPRE}. Thus when $l_b/l_c$ becomes very small, by fixing $\lambda_\mathrm{b} = l_c$, one obtains $\zeta = 1$ and $\lambda_\mathrm{NA} = l_c^2/l_b$. Since the non-affinity length scale obtained under the assumption of $\lambda_\mathrm{b} = l_c$ is the same as found in lattice based 2D and 3D networks, it seems that indeed, the bending correlation length is very close to $l_c$. One does not expect this to hold for $L$ approaching $L_\mathrm{r}$ where highly non-affine deformations would include bending that occurs on length scales much larger than $l_c$. However, as we show above, by making an empirical correction to the length, i.e., replacing $L$ by $(L-L_\mathrm{r})$, the scaling Eq.~\eqref{eqn:linmodscaling} is extended all the way up to the minimum length $L_\mathrm{r}$ required for rigidity percolation.

As shown above, the primary difference between the two types of network structures, lattice and off-lattice, is in their bending correlation length. However, with appropriately chosen exponent $\zeta$, the linear modulus from both off-lattice and lattice based networks can be quantitatively captured by Eq.~\eqref{eqn:linmodscaling}. Thus, we conclude that Eqs.~\eqref{eqn:zeta} and \eqref{eqn:linmodscaling} give a unified description of the linear mechanics of fibrous networks independent of the detailed microstructure. In the next section, we focus on the stiffening regime, $\gamma_0 \leq \gamma \leq \gamma_\mathrm{c}$. We demonstrate that independent of the details of the network, the nonlinear mechanics can also be described in a unified way.

\section{Nonlinear Regime}
\label{sec:nonlinregime}
The shear and normal stress are shown in Fig.~\ref{fig:Svgamma}. In the linear regime, $\sigma$ is linear in strain while $\tau$ is always negative and quadratic as expected from symmetry arguments~\cite{art:Poynting,art:Janmey2,art:HeussingerPRE07,art:Conti}. The negative sign in the normal stress is characteristic of biopolymer gels and has been observed in experiments~\cite{art:Janmey2}, where it was attributed to the asymmetric thermal force-extension curve of the constituent fibers \cite{art:MacKintosh95} or to non-affine deformations of athermal networks~\cite{art:hatami2008scaling, art:van2008models,art:basu2011nonaffine}, which lead to an effective network-level asymmetry in the response~\cite{art:HeussingerPRE07,art:Conti}. For very low strains, $\sigma\sim\gamma$ and $-\tau\sim\gamma^2$. As $\gamma$ increases, the shear and normal stress become increasingly comparable in magnitude. We define $\gamma_0$ as the strain at which $|\sigma|=|\tau|$, above which both stresses rapidly increase as the strain approaches $\gamma_\mathrm{c}$. For $\gamma>\gamma_\mathrm{c}$, both stress curves converge to their respective $\kappa=0$ limits similarly observed for the $K$ vs $\gamma$ curves in Fig.~\ref{fig:Kvgamma}. In the large strain limit, the shear response is again linear in strain, while the normal response approaches a constant.

An interesting feature of the strain stiffening regime can be observed in the $K$ vs $\sigma$ curves shown in Fig. ~\ref{fig:KvS}, which reveals two distinct nonlinear stiffening regimes: a bend-dominated stiffening initiated by the points $(\sigma_0,K_0)$ at the onset strain $\gamma_0$ which proceeds to stiffen as $K\sim\sigma^{\alpha}$, with $\alpha$ increasing for decreasing $\tilde{\kappa}$ (lower inset of Fig.~\ref{fig:KvS}); and a stretch-dominated stiffening where all curves converge to $K\sim\sigma^{1/2}$~\cite{art:Conti,art:BroederszSM2011,art:buxton2007bending}. These results are consistent with prior theoretical work showing an evolution of exponents from $\alpha\simeq 1/2$ through $\alpha\simeq 1$ and higher values with decreasing $\tilde\kappa$~\cite{art:BroederszSM2011}. Such an evolution of the stiffening exponent with fiber rigidity is also consistent with recent experiments on collagen networks~\cite{art:LicupPNAS}.
In contrast to what has been proposed in~\cite{art:Zagar,art:zagar2011elasticity}, however, our results show that there is no unique exponent $\alpha=3/2$ in the initial stiffening regime. 

\subsection{Onset of strain stiffening}
\label{subsec:stiffening}
As mentioned above, the strain $\gamma_0$ at the onset of stiffening is characterized by the points of stiffness $K_0$ scaling linearly with shear stress $\sigma_0$. This feature can be understood as follows. At low stresses, the elastic energy density is dominated by soft bending modes and therefore $G\sim\tilde{\kappa}$ (Fig. \ref{fig:Kvgamma}, inset)~\cite{art:Kroy,art:Satcher}. Moreover, these networks stiffen at an onset stress $\sigma_0$ proportional to $\tilde{\kappa}$ (Fig. \ref{fig:Svgamma}, inset), which coincides with the onset of fiber buckling~\cite{art:Conti,art:Onck}. From these observations, together with the fact that $K$ and $\sigma$ have the same units, it follows that $K_0\approx G$ and $\sigma_0$ should depend in the same way on network parameters. Thus, the points $(\sigma_0,K_0)$ should exhibit a linear relationship, as seen in networks for $\tilde{\kappa}\lesssim 10^{-2}$, which means that in these bend-dominated networks, the onset strain $\gamma_0\sim\sigma_0/K_0$ is independent of $\tilde{\kappa}$ (inset, Fig. \ref{fig:KvS}). The independence of $\gamma_0$ on material parameters such as fiber rigidity or concentration suggests that there is no intrinsic length scale besides $l_c$ that governs the response in the stiffening regime. This $\tilde{\kappa}$-independent regime is fully describable by a network of floppy rope-like fibers, and can be captured by our $\tilde{\kappa}=0$ limit. In what follows, we will first derive the onset of nonlinear stiffening in this limit using pure geometric relaxation arguments to obtain $\gamma_0\rightarrow\gamma_\mathrm{g}$. We then build up from this result to obtain a generalized $\gamma_0$ for networks of finite $\tilde{\kappa}$.
\begin{figure}[t]
\centering
\includegraphics[width=0.47\textwidth]{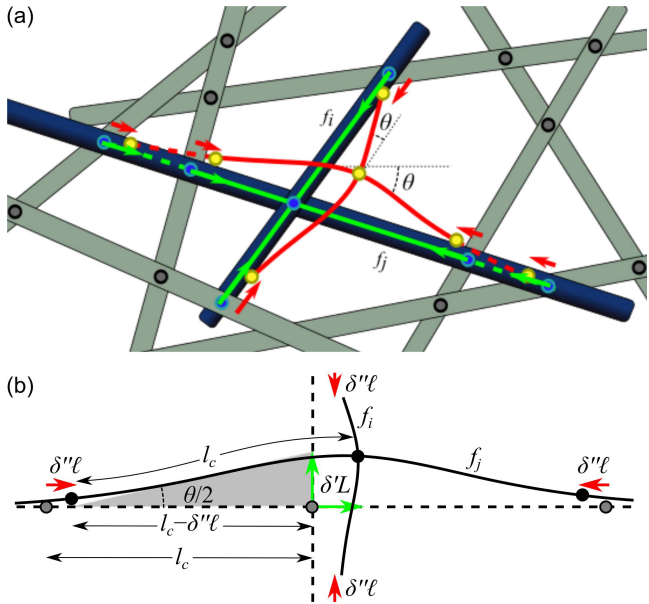}
\caption{(Color online) (a) Schematic showing two interacting fiber strands $f_i$ and $f_j$ before (green) and after (red) relaxation. Circles denote points of mechanical constraints also shown before (blue) and after (yellow) relaxation. The backbone relaxation $\delta'L$ of $f_i$ (green arrows) induces bending angles $\theta$ and longitudinal displacements $\delta''\ell$ (red arrows) on $f_j$, and vice versa. (b) A simplified diagram of the interacting strands before (dashed lines) and after (solid curves) relaxation shows the geometric relation between the coupled displacements $\delta'L$ and $\delta''\ell$ (gray triangle).}
\label{fig:schematic}
\end{figure}

Stiffening should therefore be understood in purely geometric terms as follows. In a network with bend-dominated linear elastic response, any fiber can relax its stored stretching energy by inducing bend amplitudes to the fiber strands directly connected to it (Fig. \ref{fig:schematic}). When a strand $f_i$ undergoes a backbone relaxation $\gamma L$, it induces on strand $f_j$ a transverse displacement $\delta'L\sim\gamma L$ and a longitudinal displacement (i.e., end-to-end contraction) $\delta''\ell$, both related as $\delta''\ell\approx\delta'L^2/l_c$ for small relaxations. These displacements are coupled since the longitudinal contraction of $f_j$ relaxes the stretching energy which it would have acquired from the transverse bending displacement. Similarly, the backbone relaxation of $f_j$ induces the same coupled displacements on $f_i$. To a first approximation, the total contraction of a fiber is given by the sum $\delta''L=\sum_{l_c}^L\delta''\ell\approx\left(\tfrac{L}{l_c}\right)\delta''\ell\sim\gamma^2 L^3/l_c^2$. For an isotropic network, the maximum strain $\gamma_\mathrm{g}$ at which the displacements are purely governed by these \emph{geometric} relaxations is when $\delta''L\approx\delta'L$. This maximum strain sets the onset of stiffening for floppy networks:
\begin{equation}
\label{eqn:gamma_g}
\gamma_0\mathop{\longrightarrow}_{\tilde{\kappa}\rightarrow 0^+} A\left(\frac{l_c}{L}\right)^2\equiv\gamma_\mathrm{g},
\end{equation}
where $A\approx 1$. This result shows that the onset of stiffening in floppy networks is determined by the cross-linking density $L/l_c$. Indeed, if there are on average few mechanical constraints attached to a fiber, the network can be deformed over a greater range where geometric relaxations can be explored.

In the linear regime where fiber relaxations mainly induce bending displacements $\theta\sim\delta'L/l_c$, the elastic energy of the network should be dominated by fiber bending $\mathcal{H}^{(\mathrm{b})}_0\sim\tfrac{\kappa}{l_c}(\tfrac{\delta'L}{l_c})^2$. However, we have seen from the above geometric picture that longitudinal displacements $\delta''L$ couple to the transverse displacements. This higher order contribution to the bending displacement is taken into account as $\theta\sim\tfrac{\delta'L}{l_c}+\tfrac{\delta''L}{l_c}$ such that
\begin{equation}
\label{eqn:Ebend}
\mathcal{H}^{(\mathrm{b})}_0\sim\frac{\kappa}{l_c}\left(\frac{\gamma L}{l_c}+\frac{\gamma^2 L^3}{l_c^3}\right)^2.
\end{equation}
One recovers Eq.~\eqref{eqn:gamma_g} when higher order contributions to $\theta$ become significant. This suggests that the onset of stiffening $\gamma_0$ is not characterized by the dominance of stretching modes in the total energy. This is in contrast to earlier studies in which the onset of nonlinearity was attributed to a transition from bending- to stretching-dominated behavior~\cite{art:Onck}.
\begin{figure}[t]
\centering
\includegraphics[width=0.48\textwidth]{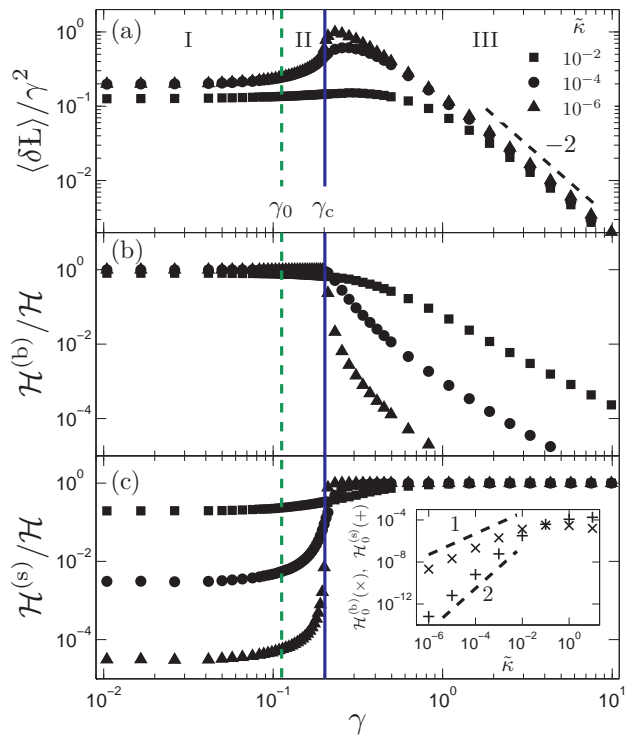}
\caption{(Color online) (a) Average fiber excess length normalized by $\gamma^2$ vs strain. The linear regime and two stiffening regimes are indicated as I, II and III, respectively. (b) Relative contributions of bending energy to the total elastic energy of the network vs strain and fiber rigidity. In regimes I and II, the total energy is dominated by bending contributions. (c) Stretching contributions become important only at strains $\gamma>\gamma_\mathrm{c}$ (III). Inset: In the linear regime, $\mathcal{H}^{(\mathrm{b})}_0\sim\tilde{\kappa}$ in networks with bend-dominated linear elasticity while $\mathcal{H}^{(\mathrm{s})}_0$ shows a quadratic $\tilde{\kappa}$-dependence.}
\label{fig:energy_ratio_exlen}
\end{figure}

The contribution of higher order bending amplitudes should correspond to a rapid increase in \emph{excess lengths}, so-called because it is a length over which one can pull an undulated fiber without stretching its backbone. For a fiber strand with contour length $l_c$ and local end-to-end length $l$ (i.e., distance between adjacent cross-links), we define the excess length as
\begin{equation}
\label{eqn:exlen}
\delta\L=\begin{cases}\delta''\ell\sim\delta'L^2/l_c, & l<l_c\\ 0, & l\geq l_c\end{cases}.
\end{equation}
As bending amplitudes develop on the strands with increasing $\gamma$, excess lengths build up as $\gamma^2$ in the linear regime. We have verified this from our simulations (Fig. \ref{fig:energy_ratio_exlen}a). Indeed, the linear regime (I) shows $\langle\delta\L\rangle/\gamma^2\sim\mathrm{const}$, followed by a rapid build-up near $\gamma_0$ (II) which peaks at $\gamma_\mathrm{c}$. For $\gamma\gg\gamma_\mathrm{c}$, the average excess length saturates to a constant (III), as one might expect for a network of stretched fibers.

\begin{figure}[t]
\centering
\includegraphics[width=0.48\textwidth]{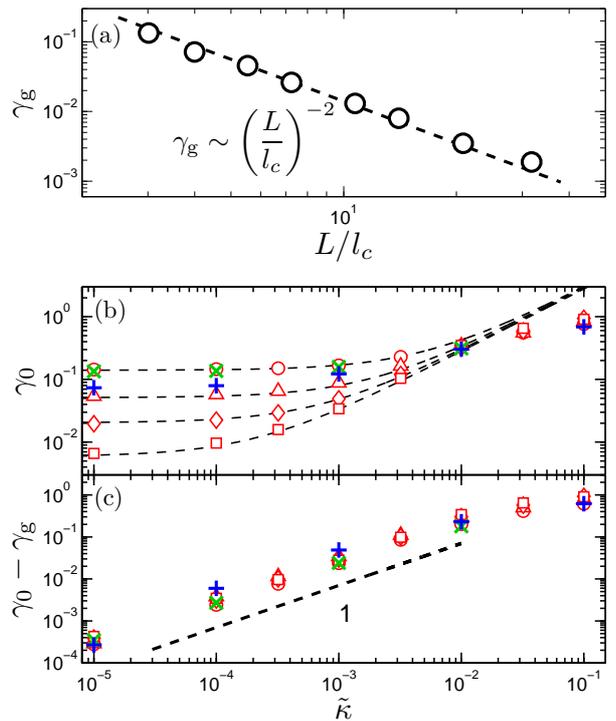}
\caption{((Color online) (a) The onset of nonlinear stiffening of a floppy network with $l_b/l_c \ll l_c/L$ shifts to lower strains with increasing $L/l_c$ as predicted by Eq.~\eqref{eqn:gamma_g}. (b) Fiber rigidity dependence of (i) $\gamma_0$ for different $L/l_c$ in a 2D lattice: $L/l_c=3$ ($\mathlarger{\mathlarger{\mathlarger{\circ}}}$), $L/l_c=6$ ($\triangle$), $L/l_c=9$ ($\Diamond$), $L/l_c=15$ ($\square$). Also shown for comparison are results from a 3D lattice ($\times$) with $L/l_c\approx 3$ and 2D Mikado ($+$) with $L/l_c\approx 11$. The dashed curves are a fit of Eq.~\eqref{eqn:gamma_o}. The constant level in the limit of $\tilde{\kappa}\rightarrow 0$ shows the value of $\gamma_\mathrm{g}$ predicted by Eq.~\eqref{eqn:gamma_g}. The onset strain $\gamma_0$ subsequently increases linearly with increasing $\tilde{\kappa}$ according to Eq.~\eqref{eqn:gamma_o}. (c) Collapse of the data from the upper panel using Eq.~\eqref{eqn:gamma_o}.}
\label{fig:onset}
\end{figure}

The relative contributions of bending and stretching energy to the total elastic energy are shown in Figs.\ \ref{fig:energy_ratio_exlen}b and \ref{fig:energy_ratio_exlen}c. As can be seen in both the linear (I) and the first stiffening (II) regimes, the total energy is dominated by fiber bending. We assume that any remaining stretching energy in a fiber strand should scale as $\mathcal{H}^{(\mathrm{s})}_0\sim\mu l_c\epsilon_r^2$ in the linear regime, where $\epsilon_r$ is some small residual strain which we shall now determine self-consistently. The bending energy in the linear regime scales accordingly as $\mathcal{H}^{(\mathrm{b})}_0\sim\tfrac{\kappa}{l_c}\left(\tfrac{(\gamma-\epsilon_r)L}{l_c}\right)^2$. Minimizing the total energy, we obtain $\epsilon_r=\tfrac{\mathcal{L}^2}{1+\mathcal{L}^2}\gamma\approx\gamma\mathcal{L}^2$, where $\mathcal{L}\equiv(l_b L)/l_c^2\ll 1$ for floppy networks. The stretching and bending energies stored in the fiber strand can now be obtained in the linear regime to leading order as:
\begin{align}
\label{eqn:Estretch_floppy}
\mathcal{H}^{(\mathrm{s})}_0&\sim\mu l_c\epsilon_r^2\approx\frac{\kappa^2 L^4}{\mu l_c^7}\gamma^2,\\
\label{eqn:Ebend_floppy}
\mathcal{H}^{(\mathrm{b})}_0&\sim\frac{\kappa}{l_c}\left(\frac{(\gamma-\epsilon_r) L}{l_c}+\frac{(\gamma-\epsilon_r)^2 L^3}{l_c^3}\right)^2\approx\frac{\kappa L^2}{l_c^3}\gamma^2.
\end{align}
Both energy contributions scale quadratically with strain in the linear regime and is confirmed by our simulations (Figs. \ref{fig:energy_ratio_exlen}b and \ref{fig:energy_ratio_exlen}c). Furthermore, the stretching contribution in floppy networks is highly suppressed because of the strong $\kappa^2$-dependence (inset, Fig. \ref{fig:energy_ratio_exlen}c). This is in contrast to what has been pointed out in a previous study~\cite{art:BroederszPRL2012} that $\mathcal{H}^{(\mathrm{s})}_0\sim\mu\gamma^4 L^4/l_c^3$. In the case of networks with finite fiber rigidity, then Eq.~\eqref{eqn:Ebend_floppy} dictates that at the onset of stiffening $\gamma=\gamma_0$, when the higher order bending term becomes comparable to the linear term, we have
\begin{equation}
\label{eqn:gamma_o}
\gamma_0\simeq\gamma_\mathrm{g}+B\tilde{\kappa},
\end{equation}
where $B\approx 28$. In the asymptotic floppy network limit where $\tilde{\kappa}\rightarrow 0$, the onset of stiffening $\gamma_0$ is determined purely by $\gamma_\mathrm{g}$ (Eq.~\eqref{eqn:gamma_g}) as shown in Fig.~\ref{fig:onset}a. This floppy limit is indicated by the finite range in $\tilde{\kappa}$ over which $\gamma_0$ is constant (Fig.~\ref{fig:onset}b). Indeed, the data from networks with different $L/l_c$ shows a good collapse of Eq.~\eqref{eqn:gamma_o} (Fig.~\ref{fig:onset}c). We note here that for large values of $\tilde{\kappa}$, the onset of nonlinearity should be dictated by the affine limit at which such rigid fibers are aligned with a $45^\circ$ angle corresponding to $100\%$ strain. Indeed, Figs.~\ref{fig:onset}b and \ref{fig:onset}c show that the onset of stiffening in networks of rigid rods saturate to $\gamma_0\rightarrow 1$.
\begin{figure}[t]
\centering
\includegraphics[width=0.47\textwidth]{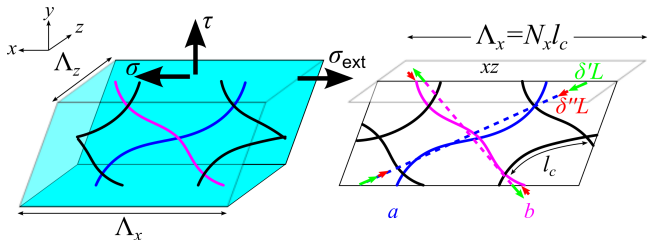}
\caption{(Color online) On the left is a schematic of the sample with the shear $\sigma$ and normal $\tau$ stresses acting on the $xz$ plane. Bold arrows indicate the stresses. The coordinates are chosen such that the internal shear stress in response to the applied external shear stress $\sigma_\mathrm{ext}$ is positive. On the cutaway view shown at the right, the dashed lines represent fibers before relaxation while solid curves represent the fibers after they have undergone the coupled relaxations $\delta'L$ (green arrows) and $\delta''L$ (red arrows). The lateral sample dimensions $\Lambda_x$ and $\Lambda_z$ can be expressed in terms of the periodicity $N_x$ and $N_z$ of fiber segments with typical spacing $l_c$.}
\label{fig:schematic1}
\end{figure}

\subsection{Stress-controlled stiffening}
\label{subsec:bend_stiff}
Three key points that characterize network stiffening in regime II of Fig.\ \ref{fig:energy_ratio_exlen} are: (i) bending modes still dominate fiber stretching since the onset of nonlinearity is not a bend-stretch transition, (ii) nonlinear buildup of excess lengths, and (iii) normal stress is negative and comparable in magnitude to the shear stress. To understand the latter, consider the mean-field representation of the network in Fig.~\ref{fig:schematic1}. Treating the fibers as bendable rods, every rod exerts a force of magnitude $F\propto\mu\epsilon_r$ on an arbitrary $xz$ plane parallel to the shear boundary. In the floppy network limit, the forces parallel and normal to the plane are $F_\parallel\sim\frac{\kappa L}{l_c^4}(\delta'L + \delta''L + \cdots)_\parallel$ and $F_\perp\sim\frac{\kappa L}{l_c^4}(\delta'L + \delta''L + \cdots)_\perp$, where other higher order relaxations can be taken into account. The contribution from the connected fiber segments $a$ and $b$ to the shear and normal stresses are respectively $\sigma\approx(F_a+F_b)_\parallel/l_c^{d-1}$ and $\tau\approx(F_a+F_b)_\perp/l_c^{d-1}$ (see Appendix). Using the expressions for the force components including higher order corrections (see Appendix) and taking into account the appropriate signs relative to the coordinate system shown in Fig.~\ref{fig:schematic1}, we have
\begin{align}
\label{eqn:Stress}
\sigma&\approx\frac{\kappa L^2}{l_c^{d+3}}\gamma+\left(\frac{L}{l_c}\right)^2\frac{\kappa L^4}{l_c^{d+5}}\gamma^3,\\
\label{eqn:Nstress}
\tau&\approx-\frac{\kappa L^4}{l_c^{d+5}}\gamma^2-\left(\frac{L}{l_c}\right)^2\frac{\kappa L^6}{l_c^{d+7}}\gamma^4.
\end{align}

\begin{figure}[t]
\centering
\includegraphics[width=0.48\textwidth]{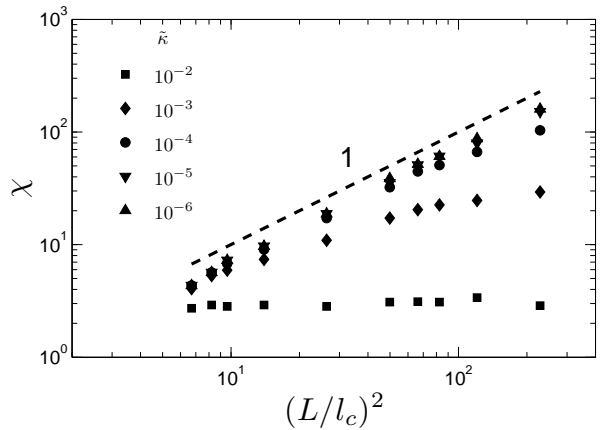}
\caption{Susceptibility to the normal stress in the nonlinear regime. As the fiber rigidity decreases, Eq.~\eqref{eqn:susceptibility} is valid for increasingly larger range of $L/l_c$.}
\label{fig:susceptibility}
\end{figure}

\noindent Thus, for floppy networks at the onset of nonlinearity (i.e., $\gamma=\gamma_0\simeq\gamma_\mathrm{g}$) we obtain the result that $\sigma\approx|\tau|\sim\kappa/l_c^{d+1}$. Furthermore, taking $K=\tfrac{\partial\sigma}{\partial\gamma}$ in combination with $|\tau|$, we obtain the stiffening relation~\cite{art:LicupPNAS}:
\begin{equation}
\label{eqn:K_hyp}
K\simeq G+\chi|\tau|,
\end{equation}
with linear modulus $G=\kappa L^2/l_c^{d+3}$ and the \emph{susceptibility}
\begin{equation}
\label{eqn:susceptibility}
\chi=(L/l_c)^2\sim\gamma_0^{-1}.
\end{equation}
This indicates that the stiffness is dominated by $G\sim\kappa$ in the linear regime while the normal stresses provide additional stabilization in the nonlinear regime. Figure \ref{fig:susceptibility} shows the susceptibility to the normal stress as a function of the cross-linking density and fiber rigidity. The floppy network limit clearly shows the relation $\chi\sim \gamma_0^{-1}$.
\begin{figure*}[t]
\subfloat[]{
  \includegraphics[width=0.47\textwidth]{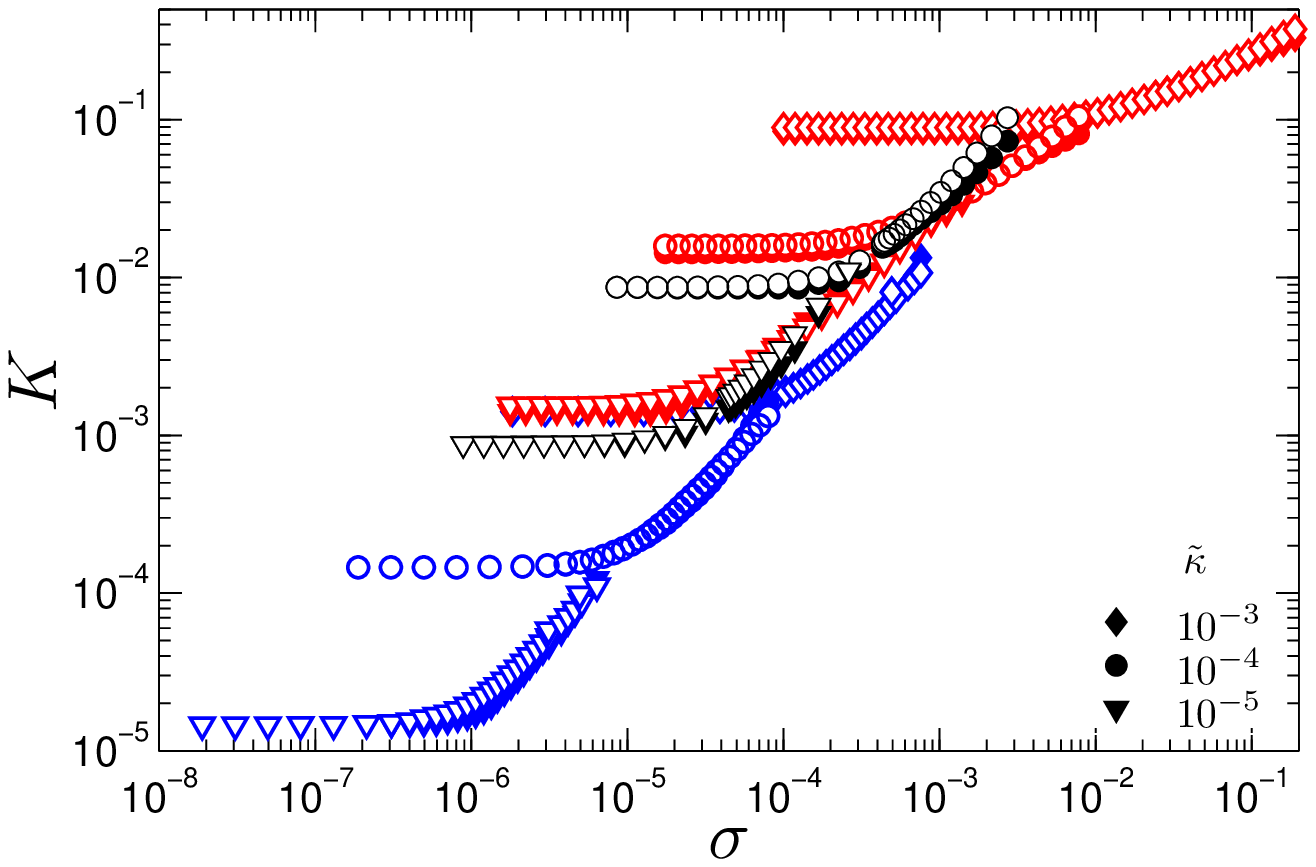}
  \label{fig:normalstiffening}
}
\qquad
\subfloat[]{
  \includegraphics[width=0.47\textwidth]{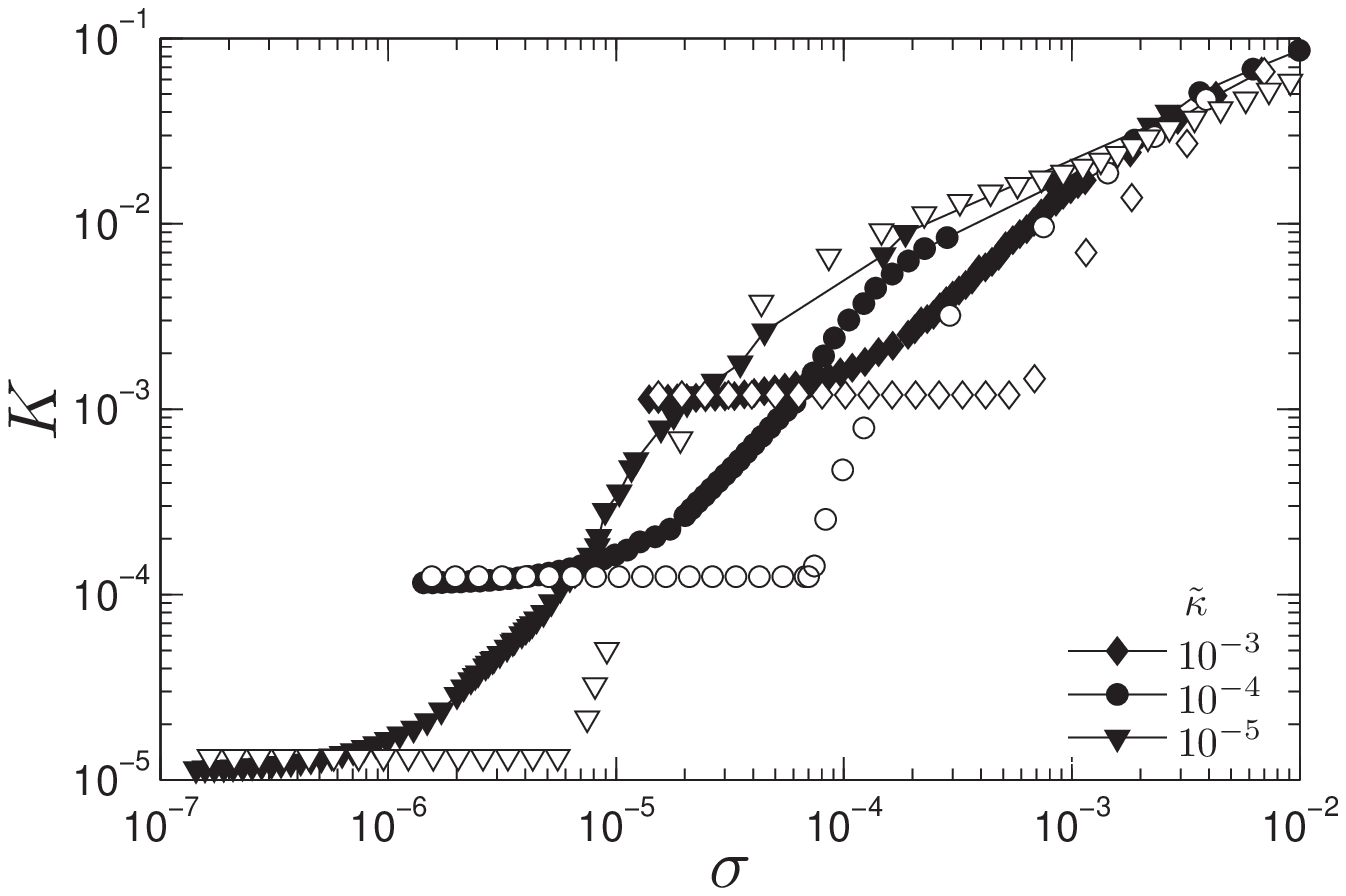}
  \label{fig:relax}
}
\caption{((Color online) (a) Stiffness vs shear stress (filled symbols) for a network with $L/l_c=3$ (blue) and $L/l_c=9$ (red) compared with the the stiffening hypothesis in Eq.~\eqref{eqn:K_hyp} (empty symbols) show that normal stresses stabilize the bend-dominated nonlinear regime. Results from 2D Mikado simulations with $L/l_c=11$ are shown in black. (b) Stiffness vs shear stress (filled symbols) for a network with $L/l_c=3$. When the shear boundaries are relaxed, the stiffness drops to the level indicated by the linear regime (open symbols).}
\end{figure*}

To test the stiffening relation in Eq.~\eqref{eqn:K_hyp}, we compare $K$ with $G+\chi|\tau|$ and plot them with $\sigma$ shown in Fig.~\ref{fig:normalstiffening}. Indeed, the linear regime is characterized by $G\sim\tilde{\kappa}$ where the magnitude of the normal stresses are not significant compared to the shear stresses. In the stiffening regime, there is excellent agreement between $K$ and $G+\chi|\tau|$. As can be seen in Fig.~\ref{fig:normalstiffening} data from mikado network also follows the stiffening relation in Eq.~\eqref{eqn:K_hyp}. As in the case of lattice based networks, the susceptibility of off-lattice networks to normal stress is the inverse of the stiffening strain. Since the stiffening strain depends on the network architecture, it appears that the stiffening relation in Eq.~\eqref{eqn:K_hyp} together with a network architecture-dependent susceptibility is a general relation to describe the nonlinear stiffening of disordered elastic networks. As a final confirmation, we perform an additional relaxation of the networks by releasing the normal stresses. Indeed, when we relax the normal stresses, the stiffness drops to the level indicated by linear modulus (Fig.~\ref{fig:relax}). This is a clear indication that the normal stresses control the nonlinear stiffening of these networks.
Moreover, the onset of stiffening with free normal boundaries occurs near $\gamma_\mathrm{c}$ at the beginning of regime III in Fig.\ \ref{fig:energy_ratio_exlen}, which is also the regime where stretching dominates, as shown in that figure. Importantly, throughout the stiffening in regime II, the bending energy still dominates the stretching energy.

\section{Discussion}
\label{sec:discussion}
Here, we have studied the elastic behavior of sub-isostatic athermal fiber networks. Athermal fiber networks can be used to model the mechanics of biological networks such as collagen. It is a priori not clear whether one needs to take into account the detailed microstructure of a biological network in a computational model to capture the mechanics. Most of the computational studies are based on lattice based~\cite{art:BroederszNatPhys,art:BroederszSM2011,art:BroederszPRL2012,art:Heussinger,art:MaoPRE042602_2013} or an off-lattice based network structures~\cite{art:Wilhelm,art:HeadPRL,art:HeadPRE,art:Conti,art:Onck,art:Huisman}. The primary advantage of a lattice based approach is the computational efficiency. By contrast, off-lattice networks, though computationally intensive, would appear to be more realistic, in the sense that the network structure has built in spatial disorder that is a key feature of biologically relevant networks. Here we show that despite the structural differences, these two approaches can be unified and are equally suited to describe most aspects of the mechanical response of athermal fiber networks. We show that the elastic modulus in the linear regime, for both lattice and off-lattice based networks, can be fully characterized in terms of a non-affinity length scale $\lambda_\mathrm{NA}$~\cite{art:HeadPRL,art:HeadPRE,art:BroederszPRL2012}, which depends on the underlying network structure. The scaling relation in Eq.~\eqref{eqn:linmodscaling} with the network-dependent exponent $\zeta$ captures the crossover behavior of the linear modulus of a network. The non-affinity length scale can be derived for a given filamentous network using mean-field arguments~\cite{art:HeadPRL,art:HeadPRE}. However, we show that with an empirical correction, replacing the filament length $L$ by $L-L_\mathrm{r}$, the scaling relation Eq.~\eqref{eqn:linmodscaling} can even capture the linear mechanics of networks close to the rigidity percolation where non mean-field behavior is expected. Our computational approach is based on networks which are composed of discrete filaments allowing for an unambiguous and intuitive definition of the non-affinity length scale $\lambda_\mathrm{NA}$. However, the concept of the non-affinity length scale can be extended to branched networks by considering the average branching distance.

\begin{figure}[b]
\centering
\includegraphics[width=0.48\textwidth]{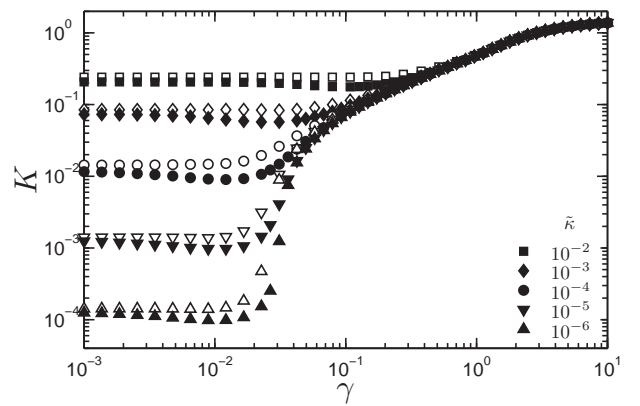}
\caption{For networks with high connectivity and cross-linking density such as in a 2D lattice with $z=3.8$ and $L/l_c=9$ (filled symbols are the red data set from Fig.~\ref{fig:Kvgamma}a), an apparent ``softening'' of the network is observed as $K$ dips slightly relative to $G$. This artifact is not present for lower $z$ and $L/l_c$ (black data set in Fig.~\ref{fig:Kvgamma}a) or when undulations are introduced to the fibers (open symbols) by applying a small uniform macroscopic compressive strain ($\varepsilon<1\%$) normal to the network boundaries.}
\label{fig:artifact}
\end{figure}

Previous computational studies on both lattice and off-lattice based networks have reported that the transition from linear to nonlinear regime under strain is marked by an initial softening of the modulus~\cite{art:Onck,art:Conti,art:abhilash2012stochastic}. The softening occurs due to buckling of the filaments under compression. However, to our knowledge, experimental demonstration of the softening has remained elusive. We suggest that the buckling-induced softening is an artifact of simulations. We show that on introducing undulations in the discrete filaments, no such softening is seen in the simulations (Fig.~\ref{fig:artifact}). Under compression, the undulating filaments undergo increased bending but do not buckle. It is expected that in any biological network, the filaments exhibit undulations, either from defects or prestress, and hence would not demonstrate buckling induced softening under strain.

The onset of the nonlinear regime is marked by a stiffening strain $\gamma_0$ at which the normal stress becomes comparable to the shear stress. We derive $\gamma_0$ using only geometric arguments and demonstrate that for bend-dominated networks, our expression is in excellent agreement with the simulations. In a bend dominated network, with increasing strain, the bend amplitude increases. The increase in bend amplitude is coupled to the longitudinal contraction of the filament along its backbone. When these two displacements, namely the contraction along the backbone and the bend amplitude become comparable, nonlinear stiffening sets in such that any further strain induces stretching of filaments in addition to bending. We also demonstrate that the above geometric argument immediately leads to normal stress becoming comparable to the shear stress at $\gamma_0$. Obtaining $\gamma_0$ as the strain at which normal and shear stress become equal provides an unambiguous definition of the onset of stiffening. Our derivation of $\gamma_0$ is purely geometrical and can be considered to hold only in the limit of vanishing bending rigidity. We derive an expression for the stiffening strain for finite bending rigidity and show that it can accurately describe the onset of stiffening for even those networks which are not bend-dominated in the linear regime. The onset of stiffening strain, as expected, reduces to $\gamma_0$ in the limit of vanishing bending rigidity. Experimental determination of $\gamma_0$ is based on an arbitrary criterion such as the strain at which the differential modulus becomes 3 times the linear modulus~\cite{art:LicupPNAS}. However, the advantage of defining $\gamma_0$ based on stress could be nullified in experiments due to \sn{the ambiguity in determining the} normal stress. Any prestress in the network would offset the normal stresses generated in the network under strain. 

In the nonlinear regime, for $\gamma>\gamma_0$ both bending and stretching energies increase faster than a quadratic dependence on the strain which manifests itself in a rapid increase in the modulus with strain. At a certain strain $\gamma_\mathrm{c} > \gamma_0$, the two energies become comparable to each other. The nonlinear mechanics in the range $\gamma_0\leq\gamma\leq\gamma_\mathrm{c}$ are controlled by normal stress in the network. We show that the elastic modulus increases in proportion to the normal stress. The observation that the modulus scales linearly with the normal stress is reminiscent of the stabilization of floppy networks under normal stress. Fiber networks, in absence of bending interactions, are floppy and can be stabilized by several fields~\cite{art:Wyart,art:BroederszNatPhys,art:DennisonPRL,art:SheinmanPRL,art:shokef2012scaling,art:Sharma,art:Fengarxiv2015} including normal stress. 
The normal stress can be generated internally by molecular motors~\cite{art:SheinmanPRL,art:shokef2012scaling} or externally by subjecting network to a global deformation~\cite{art:Heidemann,art:amuasi2015nonlinear}. 
Independent of the origin of the normal stress, the linear modulus of an initially unstable network (in absence of normal stress) scales linearly with the normal stress. Here, we generalize the idea of stabilization by normal stress to an initially stable network (finite bending interactions) in the nonlinear regime, where the normal stress become the dominant stress in the network and control the stiffening. We present a scaling argument which yields a linear relation between the nonlinear modulus and the normal stress in the stiffening regime. The modulus and the normal stress are related via the network susceptibility to the latter. We show that the susceptibility is fully governed by the underlying geometry of the network. In fact, the susceptibility scales as the inverse $\gamma_0$. To further test the role of normal stress in stiffening regime, we consider a scenario in which normal stress is always relaxed to zero for any imposed shear strain by allowing the shear boundaries to retract along the normal direction. We observe that there is no stiffening in the absence of normal stress. The modulus remains clamped to the linear modulus in the regime $\gamma_0 \leq \gamma < \gamma_\mathrm{c}$. Expriments on collagen networks have shown that over a wide range of collagen concentration, $K$ scales linearly with the shear stress $\sigma$~\cite{art:LicupPNAS,art:fung1967elasticity}. We show that such dependence of $K$ on the shear stress follows naturally from our hypothesis of normal stress induced stiffening. Over a significant range of bending rigidity which is directly related to protein concentration~\cite{art:LicupPNAS}, we find that the shear stress scales approximately linearly with the normal stress. It follows that stiffening can be understood in terms of normal stresses.


In summary, we study the mechanics of athermal fiber networks. The linear mechanics can be captured in terms of non-affinity length scale. The nonlinear mechanics can be considered as composed of two regimes. From the onset of stiffening to a critical strain, the first regime, the stiffening is governed by strain-induced normal stresses. Beyond the critical strain, the stiffening is governed by stretching of filaments. Our study provides a general framework to capture linear and nonlinear mechanics of fiber networks for both lattice and off-lattice based network structures.

\appendix*
\section{}
\subsection*{Line density calculation of lattice-based networks}
\label{apx:linedensity}
On any lattice with uniform bond lengths $l_c$, the line density can be calculated as the total length of bonds per unit volume, i.e., $\rho=n_bl_c/v_0$ where $n_s$ is the number of bonds in a unit cell of volume $v_0$. In a two-dimensional diluted triangular lattice, a unit cell has each bond shared by two triangles, so that $n_s=\frac{3}{2}p$, where $p$ is the probability that a bond exists. With $v_0=\frac{\oldsqrt{3}}{4}l_c^2$, we obtain \[\rho_\mathrm{2D}=\frac{\frac{3}{2}pl_c}{\frac{\oldsqrt{3}}{4}l_c^2}=\frac{\tilde{\rho}_\mathrm{2D}}{l_c};\enskip\tilde{\rho}_\mathrm{2D}=\frac{6p}{\oldsqrt{3}}.\] In the case of a 3D diluted FCC lattice, we can imagine six lines intersect each vertex. Enclosing a vertex by a sphere of radius $l_c/2$, the total length of the enclosed bonds is $6pl_c$. Dividing by the volume of the sphere and multiplying by the packing fraction of the FCC lattice which is $\pi/\oldsqrt{18}$, we have \[\rho_\mathrm{3D}=\frac{6pl_c}{\frac{4}{3}\pi\left(\frac{l_c}{2}\right)^3}\left(\frac{\pi}{\oldsqrt{18}}\right)=\frac{\tilde{\rho}_\mathrm{3D}}{l_c^2};\enskip\tilde{\rho}_\mathrm{3D}=\frac{12p}{\oldsqrt{2}}.\]

\subsection*{Shear and normal stresses on a boundary due to connected elastic rods}
\label{apx:stress}
We use a mean-field scaling argument to derive the shear and normal stresses on the boundary of a sample under simple shear strain. Referring to Fig.~\ref{fig:schematic1}, we assume that the fiber crossings are spaced at $l_c$ and have a periodicity along the lateral boundaries $N_x$ and $N_z$. Every fiber is an elastic rod with stretch modulus $\mu$ and bending rigidity $\kappa$. Each rod exerts a force of magnitude $F\propto\mu\epsilon_r\approx\frac{\kappa L^2}{l_c^4}\gamma$. The last approximation is when we take the floppy limit for the residual stretch $\epsilon_r$. As derived in Sec.~\ref{subsec:stiffening}, the lowest order backbone relaxations are $\delta'L\sim\gamma L$ and $\delta''L\sim\gamma^2 L^3/l_c^2$, so we can express $F$ to first order as $F\sim\frac{\kappa L}{l_c^4}\delta'L$. In general if we include higher order fiber relaxations, we should be able to write
\[
F\sim\frac{\kappa L}{l_c^4}(\delta'L + \delta''L + \delta'''L + \cdots).
\]

We can calculate stresses by summing up the components parallel and perpendicular to the shear boundary of the forces due to the relaxations of the crossed fibers $a$ and $b$. We take the lateral dimensions $\Lambda_x=N_xl_c$ and $\Lambda_z=N_zl_c$. In a 3D system, the shear/normal stress is calculated by summing up the parallel/perpendicular components of $F$ along the shear boundary:
\begin{align*}
\sigma&=\frac{\sum_{i\in{x,z}} \sum_j^{N_i}(F_a+F_b)_{\parallel j}}{\prod_{i\in{x,z}}\Lambda_i}\sim\frac{N_x N_z (F_a+F_b)_\parallel}{\Lambda_x\Lambda_z}\\
&\approx(F_a+F_b)_\parallel/l_c^2,\\
\tau&=\frac{\sum_{i\in{x,z}} \sum_j^{N_i}(F_a+F_b)_{\perp j}}{\prod_{i\in{x,z}}\Lambda_i}\sim\frac{N_x N_z (F_a+F_b)_\perp}{\Lambda_x\Lambda_z}\\
&\approx(F_a+F_b)_\perp/l_c^2.
\end{align*}
In a 2D system, these should easily translate to $\sigma\approx(F_a+F_b)_\parallel/l_c$ and $\tau\approx(F_a+F_b)_\perp/l_c$. We proceed to calculate the stresses in either $d=2$ or $d=3$ systems by substituting the force components:
\begin{align*}
\sigma&\approx\frac{\kappa L}{l_c^{d+3}}[(\delta'L_a + \cancel{\delta''L_a})+(\delta'L_b - \cancel{\delta''L_b})]_\parallel\\
\tau&\approx\frac{\kappa L}{l_c^{d+3}}[(-\bcancel{\delta'L_a} - \delta''L_a)+(\bcancel{\delta'L_b} - \delta''L_b)]_\perp
\end{align*}
where the cancellation of terms come from the mean-field assumption on the relaxations leading to the result one obtains in the linear regime:
\begin{align*}
\sigma&\sim\frac{\kappa L}{l_c^{d+3}}\delta'L\approx\frac{\kappa L^2}{l_c^{d+3}}\gamma,\\
\tau&\sim-\frac{\kappa L}{l_c^{d+3}}\delta''L\approx-\frac{\kappa L^4}{l_c^{d+5}}\gamma^2.
\end{align*}
Invoking symmetry properties of $\sigma$ and $\tau$, we generalize the above as
\begin{align*}
\sigma&\sim\frac{\kappa L}{l_c^{d+3}}(\delta'L + \delta'''L + \cdots),\\
\tau&\sim-\frac{\kappa L}{l_c^{d+3}}(\delta''L + \delta^\mathrm{(iv)}L + \cdots).
\end{align*}
\vskip 1cm

\begin{figure}[t]
\centering
\includegraphics[width=0.4\textwidth]{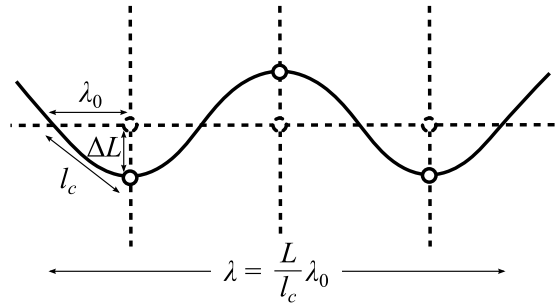}
\caption{Schematic of a fiber (dashed horizontal line) undergoing relaxation (solid curve). Other fibers are also shown with connections indicated by circles. The relaxed length is defined as $\lambda$ in terms of the relaxed segment length $\lambda_0$.}
\label{fig:schematic2}
\end{figure}
\vskip 1cm
We now obtain the higher order relaxation terms $\delta'''L$ and $\delta^\mathrm{(iv)}L$. From the diagram shown in Fig.~\ref{fig:schematic2}, we define the generalized bending amplitude $\Delta L\approx\delta'L + \delta''L$ and obtain the relaxed fiber length:
\[
\lambda=L\left[1-\left(\frac{\Delta L}{l_c}\right)^2\right]^\frac{1}{2}\approx L-\frac{\gamma^2L^3}{l_c^2}-\frac{\gamma^3L^5}{l_c^4}-\frac{\gamma^4L^7}{l_c^6}-\cdots
\]
The resulting length change of the fiber can now be written as
\begin{align*}
\Delta L&=\delta'L + \delta''L + \delta'''L + \delta^\mathrm{(iv)}L + \cdots\\
&=\gamma L+\frac{\gamma^2L^3}{l_c^2}+\frac{\gamma^3L^5}{l_c^4}+\frac{\gamma^4L^7}{l_c^6} + \cdots
\end{align*}
such that
\[
\delta'''L\sim\frac{\gamma^3L^5}{l_c^4},\enskip\delta^\mathrm{(iv)}L\sim\frac{\gamma^4L^7}{l_c^6}.
\]
Finally, we substitute these higher order relaxation terms into the generalized shear and normal stresses leading to
\begin{align*}
\sigma&\approx\frac{\kappa L^2}{l_c^{d+3}}\gamma+\left(\frac{L}{l_c}\right)^2\frac{\kappa L^4}{l_c^{d+5}}\gamma^3,\\
\tau&\approx-\frac{\kappa L^4}{l_c^{d+5}}\gamma^2-\left(\frac{L}{l_c}\right)^2\frac{\kappa L^6}{l_c^{d+7}}\gamma^4.
\end{align*}


%


\begin{thebibliography}{50}
\bibitem{art:JanmeyCellBio} P.A. Janmey, Curr. Opin. Cell Biol. \textbf{3}, 4 (1991).
\bibitem{art:wachsstock1994cross} D.H. Wachsstock, and W.H. Schwarz, and T.D. Pollard, Biophys. J. \textbf{66}, 801 (1994).
\bibitem{art:Kasza} K.E. Kasza, A.C. Rowat, J.Y. Liu, T.E. Angelini, C.P. Brangwynne, G.H. Koenderink, and D.A. Weitz, Curr. Opin. Cell Biol. \textbf{19}, 1 (2007).
\bibitem{art:Bausch} A.R. Bausch and K. Kroy, Nature Physics \textbf{2}, 4 (2006).
\bibitem{art:Fletcher} D.A. Fletcher and D. Mullins, Nature Physics \textbf{463}, 485 (2010).
\bibitem{art:chaudhuri2007reversible} O. Chaudhuri, and S.H. Parekh, and D.A. Fletcher, Nature \textbf{445}, 295 (2007).
\bibitem{art:tharmann2007viscoelasticity} R. Tharmann, and M.M.A.E. Claessens, and A.R. Bausch, Phys. Rev. Lett. \textbf{98}, 088103 (2007).
\bibitem{art:picu2011mechanics} R.C. Picu, Soft Matter \textbf{7}, 6768, (2011).
\bibitem{art:BroederszRMP} C.P. Broedersz, and F.C. MacKintosh, Rev. Mod. Phys. \textbf{86}, 995-1036 (2014).
\bibitem{art:Gardel} M.L. Gardel, J.H. Shin, F.C. MacKintosh, L. Mahadevan, P. Matsudaira, and D.A. Weitz, Science \textbf{304}, 1301 (2004).
\bibitem{art:janmey1983rheology} P.A. Janmey, and E. J. Amis, and J.D. Ferry, Journal of Rheology \textbf{27}, 135 (1983).
\bibitem{art:Xu} J. Xu, Y. Tseng, and D. Wirtz, J. Biological Chemistry \textbf{275}, 46 (2000)
\bibitem{art:Storm} C. Storm, J.J. Pastore, F.C. MacKintosh, T.C. Lubensky, and P.A. Janmey, Nature \textbf{45}, 191 (2005).
\bibitem{art:didonna2006filamin} B.A. DiDonna and A.J. Levine, Phys. Rev. Lett. \textbf{97}, 068104 (2006).
\bibitem{art:GardelPNAS06} M.L. Gardel, et al., Proc. Natl. Acad. Sci. USA \textbf{103}, 1762 (2006).
\bibitem{art:WagnerPNAS06} B. Wagner, et al., Proc. Natl. Acad. Sci. USA \textbf{103}, 13974 (2006).
\bibitem{art:kabla2007nonlinear} A. Kabla, and L. Mahadevan, Journal of The Royal Society Interface \textbf{4}, 99, (2007).
\bibitem{art:KaszaPRE09} K.E. Kasza, et al., Phys. Rev. E \textbf{79}, 041928 (2009).
\bibitem{art:stein2011micromechanics} A.M. Stein, and D.A. Vader, and D.A. Weitz, and L.M. Sander, Complexity, \textbf{16}, 22, (2011).
\bibitem{art:Saraf} H. Saraf, K.T. Ramesh, A.M. Lennon, A.C. Merkle, and J.C. Roberts, J. Biomechanics \textbf{40}, 1960 (2007).
\bibitem{art:Poynting} J.H. Poynting, Proc. R. Soc. Lond. A \textbf{82}, 546 (1909); J.H. Poynting, Proc. R. Soc. Lond. A \textbf{86}, 534 (1912).
\bibitem{art:Janmey2} P.A. Janmey, M.E. McCormick, S. Rammensee, J.L. Leight, P.C. Georges, and F.C. MacKintosh, Nature Materials \textbf{6}, 48 (2007).
\bibitem{art:HeussingerPRE07} C. Heussinger, B. Schaefer, and E. Frey, Phys. Rev. E \textbf{76}, 031906 (2007).
\bibitem{art:Conti} E. Conti and F.C. MacKintosh, Phys Rev. Lett. \textbf{102}, 088102 (2009).
\bibitem{art:Lindstrom} S.B. Lindstr\"om, D.A. Vader, A. Kulachenko, and D.A. Weitz, Phys. Rev. E \textbf{82}, 051905 (2010).
\bibitem{art:LicupPNAS} A.J. Licup, S. M\"unster, A. Sharma, M. Sheinman, L. Jawerth, B. Fabry, D.A. Weitz, and F.C. MacKintosh, Proc. Natl. Acad. Sci. USA \textbf{112}, 31 (2015).
\bibitem{art:Sharma} A. Sharma, A.J. Licup, R. Rens, M. Sheinman, K. Jansen, G. Koenderink, and F.C. MacKintosh, arXiv:1506.07792.
\bibitem{art:Wyart} M. Wyart, H. Liang, A. Kabla, and L. Mahadevan, Phys. Rev. Lett. \textbf{101}, 215501 (2008).
\bibitem{art:BroederszNatPhys} C.P. Broedersz, X. Mao, T.C. Lubensky, and F.C. MacKintosh, Nature Physics \textbf{7}, 983 (2011).
\bibitem{art:DennisonPRL} M. Dennison, M. Sheinman, C. Storm, and F.C. MacKintosh, Phys. Rev. Lett. \textbf{111}, 095503 (2013).
\bibitem{art:SheinmanPRL} M. Sheinman, C.P. Broedersz, and F.C. MacKintosh, Phys. Rev. Lett. \textbf{109}, 238101 (2012).
\bibitem{art:shokef2012scaling} Y. Shokef and S.A. Safran, Phys. Rev. Lett. \textbf{108}, 178103 (2012).
\bibitem{art:Fengarxiv2015} J. Feng, H. Levine, X. Mao, and L. M. Sander, arXiv:1507.075192.
\bibitem{art:Wilhelm} J. Wilhelm and E. Frey, Phys Rev. Lett. \textbf{91}, 108103 (2003).
\bibitem{art:HeadPRL} D.A. Head, A.J. Levine, and F.C. MacKintosh, Phys Rev. Lett. \textbf{91}, 108102 (2003).
\bibitem{art:HeadPRE} D.A. Head, A.J. Levine, and F.C. MacKintosh, Phys. Rev. E \textbf{68}, 061907 (2003).
\bibitem{art:DasPRL2007} M. Das, F.C. MacKintosh, and A.J. Levine, Phys Rev. Lett. \textbf{99}, 038101 (2007).
\bibitem{art:Onck} P.R. Onck, T. Koeman, T. van Dillen, and E. van der Giessen, Phys. Rev. Lett \textbf{95}, 178102 (2005).
\bibitem{art:Huisman} E.M. Huisman, C. Storm, and G.T. Barkema, Phys. Rev. E \textbf{78}, 051801 (2008).
\bibitem{art:shahsavari2012model} A. Shahsavari, and R.C. Picu, Phys. Rev. E \textbf{86}, 011923, (2012).
\bibitem{art:BroederszSM2011} C.P. Broedersz and F.C. MacKintosh, Soft Matter \textbf{7}, 3186 (2011).
\bibitem{art:BroederszPRL2012} C.P. Broedersz, M. Sheinman, and F.C. MacKintosh, Phys. Rev. Lett. \textbf{108}, 078102 (2012).
\bibitem{art:Heussinger} C. Heussinger and E. Frey, Phys. Rev. E \textbf{75}, 011917 (2007).
\bibitem{art:MaoPRE042602_2013} X. Mao, O. Stenull, and T.C. Lubensky, Phys. Rev. E \textbf{87}, 042602 (2013).
\bibitem{art:MaoPRE042601_2013} X. Mao, O. Stenull, and T.C. Lubensky, Phys. Rev. E \textbf{87}, 042601 (2013).
\bibitem{art:SheinmanPRE} M. Sheinman, C.P. Broedersz, and F.C. MacKintosh, Phys. Rev. E \textbf{85}, 021801 (2012).
\bibitem{art:huisman2007three} E.M. Huisman, and T. van Dillen, and P.R. Onck, and E. Van der Giessen,  Phys. Rev. Lett. \textbf{99}, 208103 (2007).
\bibitem{art:Maxwell} J.C. Maxwell, Philo. Mag. \textbf{27}, 182 (1864).
\bibitem{art:Alexander} S. Alexander, Phys. Rep. \textbf{296}, 2 (1998).
\bibitem{book:LandauLifshitz} L.D. Landau, and E.M. Lifshitz, Theory of Elasticity, 2nd ed. \emph{The Equilibrium of Rods and Plates} (Pergamon Press, Oxford), pp 44--97 (1970).
\bibitem{art:HeadPRE_R} D.A. Head, F.C. MacKintosh, and A.J. Levine, Phys. Rev. E \textbf{68}, 025101(R) (2003).
\bibitem{book:numrec} W.H. Press, S.A. Teukolsky, W.T. Vetterling, and B.P. Flannery, \emph{Numerical Recipes in C: The Art of Scientific Computing} (Cambridge University Press, 1992), 2nd ed.
\bibitem{art:LeesEdwards} A.W. Lees and S.F. Edwards, J. Phys. C Solid State Phys. \textbf{5}, 15 (1972).
\bibitem{art:Jaspers} M. Jaspers, M. Dennison, M.F.J. Mabesoone, F.C. MacKintosh, A.E. Rowan, and P.H.J. Kouwer, Nat. Comm. \textbf{5}, 5808 (2014).
\bibitem{art:MacKintosh95} F.C. MacKintosh, J. K\"as, and P.A. Janmey, Phys. Rev. Lett. \textbf{75}, 4425 (1995).
\bibitem{art:hatami2008scaling} H. Hatami-Marbini, and R.C. Picu, Phys. Rev. E \textbf{77}, 062103, (2008).
\bibitem{art:van2008models} T. van Dillen, and P.R. Onck, and E. Van der Giessen, Journal of the Mechanics and Physics of Solids, \textbf{56}, 2240, (2008).
\bibitem{art:basu2011nonaffine} A. Basu, and Q. Wen, and X. Mao, and T.C. Lubensky, and P.A. Janmey, and A.G. Yodh, Macromolecules \textbf{44}, 1671, (2011).
\bibitem{art:buxton2007bending} G.A. Buxton, and N. Clarke, Phys. Rev. Lett. \textbf{98}, 238103 (2007).
\bibitem{art:zagar2011elasticity} G. Zagar, and P.R. Onck, and E. Van der Giessen, Macromolecules \textbf{44}, 7026, (2011).
\bibitem{art:Zagar} G. \v Zagar, P.R. Onck, and E. Van der Giessen, Biophys. J. \textbf{108}, 6 (2015).
\bibitem{art:Kroy} K. Kroy, and E. Frey, Phys. Rev. Lett. \textbf{77}, 306 (1996).
\bibitem{art:Satcher} R.L. Satcher, and C.F. Dewey Jr., Biophys. J. \textbf{71}, 1 (1996).
\bibitem{art:abhilash2012stochastic} A.S. Abhilash, and P.K. Purohit, and S.P. Joshi, Soft Matter \textbf{8}, 7004 (2012).
\bibitem{art:amuasi2015nonlinear} H.E. Amuasi, and C. Heussinger, R.L.C. Vink, RLC and A. Zippelius, New J. of Phys. \textbf{17}, 083035 (2015).
\bibitem{art:Heidemann} K.M. Heidemann, A. Sharma, F. Rehfeldt, C.F. Schmidt, and M. Wardetzky, Soft Matter \textbf{11}, 343 (2015).
\bibitem{art:fung1967elasticity} Y.C. Fung, American Journal of Physiology \textbf{213}, 1532 (1967).
\end{thebibliography}
\end{document}